\documentclass[%
twocolumn, 
notitlepage, 
aps,
prb, 
10pt, 
superscriptaddress,
floatfix,
letterpaper,
reprint,
bibnotes,
amsmath,amssymb
]{revtex4-2}
 
\usepackage{hyperref}
\usepackage{graphicx}
\usepackage{dcolumn}
\usepackage{bm}
\usepackage[compat=1.1.0]{tikz-feynman}
\usepackage{physics}
\usepackage{lipsum}

\definecolor{linkcolor}{HTML}{399B03}
\definecolor{urlcolor}{HTML}{399B03}
\hypersetup{pdfstartview=FitH, linkcolor=linkcolor,urlcolor=urlcolor,colorlinks=true}
\hypersetup{
    unicode=false,          
    pdftoolbar=true,        
    pdfmenubar=true,        
    pdffitwindow=false,     
    pdfstartview={FitH},    
    pdfauthor={},         
    colorlinks=true,        
    linkcolor=blue,         
    citecolor=red,          
}

\newcommand{\dprod}[2]{\mathbf{#1} \cdot \mathbf{#2}}

\begin{document}

\title{Low-Scaling algorithms for \texorpdfstring{$GW$}{GW} and constrained random phase approximation using symmetry-adapted interpolative separable density fitting}

\author{Chia-Nan Yeh}%
\affiliation{%
Center for Computational Quantum Physics, Flatiron Institute, New York, New York 10010, USA
}%

\author{Miguel A. Morales}%
\affiliation{%
Center for Computational Quantum Physics, Flatiron Institute, New York, New York 10010, USA
}%

\date{\today}

\begin{abstract}
We present low-scaling algorithms for $GW$ and constrained random phase approximation based on a symmetry-adapted interpolative separable density fitting (ISDF) procedure that incorporates the space-group symmetries of crystalline systems. 
The resulting formulations scale cubically with respect to system sizes and linearly with the number of $\mathbf{k}$-points, regardless of the choice of single-particle basis and whether a quasiparticle approximation is employed. 
We validate these methods through comparisons with published literature and demonstrate their efficiency in treating large-scale systems through the construction of downfolded many-body Hamiltonians for carbon dimer defects embedded in hexagonal boron nitride supercells. 
Our work highlights the efficiency and general applicability of ISDF in the context of large-scale many-body calculations with $\mathbf{k}$-point sampling beyond density functional theory. 
\end{abstract}

\maketitle

\section{Introduction \label{sec:intro}}
The understanding of electronic properties in materials has been significantly advanced by recent developments in many-body perturbation theories (MBPTs)~\cite{DFTvsMBPT_Angel2002}, a powerful tool for accessing excitation spectra. Specifically, those based on Hedin's equations~\cite{Hedin65} provide a systematic and rigorous framework for describing many-body effects arising from electron-electron interactions~\cite{DFTvsMBPT_Angel2002}.
In the context of quantum embedding, the diagrammatic formulation of MBPTs enables a consistent integration with high-level quantum many-body methods, overcoming the double-counting problems that appear in density functional theory plus dynamical mean-field theory (DFT+DMFT).

 Among various approximations to Hedin's equations, the $GW$ method~\cite{Hedin65} emerges as the simplest variant for calculating the self-energy, approximating the screened interaction using the random phase approximation (RPA). 
Simultaneously, the constrained RPA (cRPA)~\cite{cRPA_Aryasetiawan2004}, which employs the same approximation to the screened interaction, stands out as the simplest variant for calculating effective interactions in low-energy models.  
However, despite their simplification, the computational demands associated with solving the equations of $GW$ and cRPA still present significant challenges, hindering their applications to large and complex systems. 
In a conventional $GW$ implementation using either a plane-waves basis or a localized basis, the evaluations of polarizability and self-energy involve summations over orbital pairs and momentum convolutions, resulting in a quartic scaling concerning system size ($N$) and a quadratic scaling with the number of $\mathbf{k}$-points for Brillouin zone (BZ) sampling ($N_{k}$).  
 
Optimizations for large-scale MBPT calculations have been extensively explored and remain an active field of research~\cite{spacetime_GW_Rojas1995,spacetime_GW_enhancement_Steinbeck2000,GW_virtuals_Bruneval2008,VASP_RPA_Kaltak2014,VASP_RPA_JCTC_Kaltak2014,sparse_sampling_Jia_2020,VASP_minimax_Kaltak2020,GW_virutals_Gao2016}. 
One notable approach is the space-time formalism~\cite{spacetime_GW_Rojas1995}, which shifts the computation of polarizability and self-energy onto spatial coordinates and the imaginary-time axis.
This leads to a more favorable cubic scaling concerning system size and a linear scaling in the number of $\mathbf{k}$-points. 
However, the efficiency of this method depends on the presence of a fast Fourier transform (FFT) grid with a moderate size, typically achieved through the use of pseudopotentials or augmentation with a localized basis for core electrons~\cite{LAPW_G0W0_basis_conv,scGW_also_vertex_Andrey_2017,scGW_VASP_2018}.

In the context of localized basis sets, such as Gaussian-type orbitals, the implementation of $GW$ and cRPA relies on decomposition schemes for electron repulsion integrals (ERIs). 
Common schemes include Cholesky decomposition (CD)~\cite{Cholesky_ERI_MOL_Beebe1977,CD_Koch2003,CD_DF_Weigend2009}, the resolution-of-identity (RI) or density fitting (DF) technique~\cite{DF_Werner2003,CD_DF_Weigend2009,DF_Ren2012,GDF_MDF_Sun2017,RSDF_HongZhou2021}, and tensor hypercontraction (THC)~\cite{THC_one_Martinez2012,LSTHC_Sherrill2012,ISDF_Lu2015,ISDF_Bloch_Lu2016,THC-RPA_CNY2023}. 
In the cases of CD and DF, low-scaling algorithms can be attained by leveraging the locality of orbitals or the sparsity of matrix elements in extensive supercell calculations~\cite{CP2K_cubicRPA_2016,lineaRPA_AO_Schurkus2016,linearRPA_Luenser2017,linearDRPA_Kallay2015}. 
However, the resulting advantageous scalings depend critically on specific electronic properties and do not extend to systems with small to intermediate sizes. 
In constrast, more aggressive decomposition schemes like THC factorize an ERI tensor into products of five second-order tensors.
This factorized representation has demonstrated its efficacy in leading to low-scaling algorithms for $GW$~\cite{TDDFT_G0W0_ISDF_Gao2020,G0W0_COHSEX_ISDF_Ma2021,separable_RI_G0W0_Blase2021} and other electronic structure methods~\cite{THC_one_Martinez2012,LSTHC_CCSD_Parrish2014,ISDF_QRCP_hybrid_Hu2017,AFQMC_ISDF_Miguel2019,THC_MP3_Joonho2020} in the context of molecules and $\Gamma$-point supercells.
Nevertheless, its application to crystalline systems with $\mathbf{k}$-point sampling is, as of now, confined to hybrid functionals~\cite{ISDF_hybridDFT_kpts_2022} and RPA correlation energy~\cite{THC-RPA_CNY2023}. 

Following the momentum-dependent interpolative separable density fitting (ISDF) procedure introduced in Ref.~\onlinecite{THC-RPA_CNY2023}, we present a symmetry-adapted extension that incorporates the space-group symmetries inherent in crystalline systems. 
The symmetry-adapted ISDF leads to a THC representation of ERIs, resulting in low-scaling algorithms for $GW$ and cRPA. 
Our symmetry-adapted approach does not depend on the assumptions of locality or sparsity and is applicable to a generic Bloch basis without the need for a pre-optimized auxiliary basis. 
Applications to higher-order MBPTs with different self-consistency types is therefore straightforward. 

The manuscript is organized as follows: In Section~\ref{sec:hedin}, we provide an overview of Hedin's equations and subsequently introduce $GW$ and cRPA as their simplified variants. 
Section~\ref{sec:bloch_sym} introduces the Bloch basis functions in the presence of space-group symmetries. Proceeding to Section~\ref{sec:isdf}, we derive the ISDF equations in the presence of space-group symmetries. 
Building upon the symmetry-adapted ISDF procedure, we then reformulate the $GW$ and cRPA equations within the symmetry-adapted ISDF formalism in Section~\ref{sec:isdf_mbpt}. 
Section~\ref{sec:comp_detail} outlines the computational details. Finally, we present the results of $GW$ and cRPA within the symmetry-adapted ISDF formalism in Section~\ref{sec:results} and conclude our findings in Section~\ref{sec:conclusion}.

\section{Hedin's equations\label{sec:hedin}}
Following Hedin's derivation in Ref.~\onlinecite{Hedin65}, a set of self-consistent equations can be formulated connecting the Green's function $G$, the screened interaction $W$, the irreducible polarizability $\Pi$, the vertex function $\Gamma$, and the self-energy $\Sigma$:  
\begin{subequations}
\begin{align}
&G(x_{1}, x_{2}) = G_{0}(x_{1}, x_{2}) + G_{0}(x_{1}, x_{3})\Sigma(x_{3}, x_{4})G(x_{4}, x_{2}) \label{eq:hedin_dysonG}\\
&W(x_{1}, x_{2}) = V(x_{1}, x_{2}) + V(x_{1}, x_{3})\Pi(x_{3}, x_{4})W(x_{4}, x_{2}) \label{eq:hedin_dysonW}\\
&\Pi(x_{1}, x_{2}) =  G(x_{1}, x_{3})\Gamma(x_{3}, x_{4}, x_{2})G(x_{4}, x_{1})\\ 
&\Gamma(x_{1}, x_{2}, x_{3}) = \delta(x_{1}, x_{3})\delta(x_{2}, x_{3}) + \label{eq:hedin_gamma}\\
& \ \ \ \ \ \ \ \ \ \ \ \ \ \ \ \frac{\delta\Sigma(x_{1},x_{2})}{\delta G(x_{3}, x_{4})}G(x_{3}, x_{5})\Gamma(x_{5}, x_{6}, x_{2})G(x_{6}, x_{4})\nonumber\\
& \Sigma(x_{1}, x_{2}) = - G(x_{1}, x_{3})\Gamma(x_{3}, x_{4}, x_{2})W(x_{4}, x_{1}).  
\end{align}
\label{eqs:hedin}
\end{subequations}
Here, $G_{0}$ is the non-interacting Green's function, $V$ is the bare Coulomb interaction, $x_{i}$ represents the space-time coordinates, and repeated arguments on the right-hand side are summed over. 
$G_{0}$ is defined with respect to the non-interacting Hamiltonian plus the Hartree potential, and thus the remaining first-order contribution in the self-energy $\Sigma$ is the Hartree-Fock (HF) exchange. 
Eq.~\ref{eq:hedin_dysonG} and~\ref{eq:hedin_dysonW} represent the Dyson equations for the interacting Green's function and the screened interaction. Among the Hedin's equations, Eq.~\ref{eq:hedin_gamma} (also called the Bethe-Saltpeter equation) is numerically the most challenging due to its unfavorable scalings in the number of orbitals and frequencies. 

In the $GW$ approximation, the vertex function $\Gamma$ is approximated to zeroth order: 
\begin{align}
\Gamma(x_{1}, x_{2}, x_{3}) = \delta(x_{1}, x_{3})\delta(x_{2}, x_{3}), 
\end{align}
resulting in simplified expressions for the irreducible polarizability and the self-energy:
\begin{align}
&\Pi(x_{1}, x_{2}) =  G(x_{1}, x_{2})G(x_{2}, x_{1}) \label{eq:hedin_gw_pi}\\ 
& \Sigma(x_{1}, x_{2}) = - G(x_{1}, x_{2})W(x_{2}, x_{1}). \label{eq:hedin_gw_sigma}
\end{align} 
Eqs.~\ref{eq:hedin_dysonG},~\ref{eq:hedin_dysonW},~\ref{eq:hedin_gw_pi}, and~\ref{eq:hedin_gw_sigma} collectively constitute the $GW$ equations, which are solved iteratively until convergence is achieved.

Despite the significant simplification compared to the original Hedin's equations, the $GW$ approximation in first-principles calculations remains computationally demanding due to the dynamic degrees of freedom, slow basis set convergence, and non-trivial self-consistency between multiple dynamic quantities. 
The state-of-the-art $GW$ algorithm is the so-called space-time formalism~\cite{spacetime_GW_Rojas1995}, summarized as follows:
\begin{subequations}
\begin{align}
&G^{\bold{k}}(\bold{r}, \bold{r}'; \tau) = \sum_{ij}\phi^{\bold{k}}_{i}(\bold{r})G^{\bold{k}}_{ij}(\tau)\phi^{\bold{k}*}_{j}(\bold{r}') \label{eq:gw_spacetime_a}\\ 
&\Pi^{\bold{R}}(\bold{r}, \bold{r}'; \tau) = G^{\bold{R}}(\bold{r}, \bold{r}';\tau) G^{-\bold{R}}(\bold{r}', \bold{r}; -\tau) \label{eq:gw_spacetime_b}\\ 
& \epsilon^{\bold{q}}_{\bold{G}\bold{G}'}(i\Omega_{n}) = \delta_{\bold{G}\bold{G}'} - \frac{\sqrt{4\pi}}{|\bold{q+G}|}\Pi^{\bold{q}}_{\bold{G}\bold{G}'}(i\Omega_{n})\frac{\sqrt{4\pi}}{|\bold{q+G'}|} \label{eq:gw_spacetime_c}\\ 
&W^{\bold{q}}_{\bold{G}\bold{G}'}(i\Omega_{n}) = \frac{\sqrt{4\pi}}{|\bold{q+G}|} \epsilon^{\bold{q},-1}_{\bold{G}\bold{G}'}(i\Omega_{n}) \frac{\sqrt{4\pi}}{|\bold{q+G'}|} \label{eq:gw_spacetime_d}\\ 
& \Sigma^{\bold{R}}(\bold{r}, \bold{r}'; \tau) = -G^{\bold{R}}(\bold{r}, \bold{r}'; \tau)W^{\bold{R}}(\bold{r}, \bold{r}'; \tau) \label{eq:gw_spacetime_e}\\
&\Sigma^{\bold{k}}_{ij}(\tau) = \sum_{\bold{G}\bold{G}'}\phi^{\bold{k}*}_{i}(\bold{G})\Sigma^{\bold{k}}_{\bold{G}\bold{G}'}(\tau)\phi^{\bold{k}}_{j}(\bold{G}'). \label{eq:gw_spacetime_f}
\end{align}
\label{eqs:gw_spacetime}
\end{subequations}
Here, the Green's function is initially projected from a single-particle Bloch basis $\{\phi^{\bold{k}}_{i}(\bold{r})\}$ to a real-space grid for the evaluation of the irreducible polarizability $\Pi^{\bold{R}}(\bold{r}, \bold{r}'; \tau)$ in space-time coordinates where $\tau\in[0,\beta]$ ($\beta$ is the inverse temperature).
The symmetric dielectric function $\epsilon^{\bold{q}}_{\bold{G}\bold{G}'}(i\Omega_{n})$ and the screened interaction $W^{\bold{q}}_{\bold{G}\bold{G}'}(i\Omega_{n})$ are then evaluated in momentum and Matsubara frequency spaces using a plane-wave basis. 
Finally, the self-energy $\Sigma^{\bold{R}}(\bold{r},\bold{r}'; \tau)$ is evaluated in space-time coordinates and transformed back to the single-particle Bloch basis. 

It is crucial to emphasize that the efficiency of the space-time formalism heavily depends on fast algorithms for Fourier transforms on the imaginary axis, as illustrated by the following equations:  
\begin{align}
&\Pi(i\Omega_{n}) = \int_{0}^{\beta}d\tau \Pi(\tau)e^{i\Omega_{n}\tau}  \label{eq:iaft_w} \\ 
&\Pi(\tau) = \frac{1}{\beta}\sum_{n=-\infty}^{\infty}\Pi(i\Omega_{n}) e^{-i\Omega_{n}\tau}, \label{eq:iaft_tau}
\end{align}
where $\Omega_{n} = 2n\pi/\beta$ ($n\in \mathbb{Z}$) are bosonic Matsubara frequencies. 
Implementing Eqs.~\ref{eq:iaft_w} and~\ref{eq:iaft_tau} straightforwardly would necessitate a large number of samplings for $\tau$ and $i\Omega_{n}$, leading to unrealistic prefactors in computational complexities and impractical memory requirements. 

Similarly, cRPA employs the same level of approximation to the vertex function $\Gamma$ and calculates the partially screened interaction $W_{r}$ for a pre-defined active subspace with a modified irreducible polarizability:
\begin{align}
\Pi_{r}(x_{1}, x_{2}) =  G_{r}(x_{1}, x_{2})G_{r}(x_{2}, x_{1}), 
\end{align}
where $G_{r}$ corresponds to the single-particle Green's function in which the projections of the active subspace are subtracted out.

\section{Bloch functions and symmetry \label{sec:bloch_sym}} 
For simplicity, let's consider a symmorphic space group \textit{G}, whose elements are defined by all rotations $\bold{S}$ of the crystal that leave its Hamiltonian $\bold{H}$ invariant, e.g. $[\bold{H},\bold{S}]=0$ 
\footnote{In this work, points in Cartesian space are represented by column vectors, while points in reciprocal space are represented by row vectors. As a result, the action of symmetry operations in Cartesian space are given by matrix-vector products, $\bold{r'}=\bold{Sr}$, while in reciprocal space the transposed operation is necessary, $\bold{G'}=\bold{GS^{T}}$. Inner products are given by: $\bold{G \cdot r}=\sum_{i}G_{i} r_{i}$ and are invariant to symmetry rotations, $\bold{G' \cdot r'} = \bold{ (GS^T) \cdot (Sr) } = \bold{G (S^{-1} S) r}  = \bold{G \cdot r}$, since rotation matrices are orthogonal, $S^T=S^{-1}$.}.  
The extension to non-symmorphic groups is straightforward. In momentum space, this implies invariance of the Hamiltonian under similarity transformations $\bold{U_S}$ associated with the elements of $\textit{G}$, $\bold{U_S} \bold{H_k} \bold{U_S}^{\dagger} = \bold{H_{kS^{-1}}}$. In real space, this implies invariance with respect to rotations of the lattice, $H(\bold{r},\bold{r'}) = H(\bold{Sr},\bold{Sr'})$. When symmetry is properly incorporated, the mean-field (MF) solutions of $\bold{H}$ will form a basis for the irreducible representations of \textit{G}, this includes Kohn-Sham (KS) DFT solutions as well as other suitable MF approximations. As a result, MF orbitals at different symmetry-related $\mathbf{k}$-points in the BZ will be related and transform according to: 
\begin{align}
\bold{U_S} \phi^{\bold{k}}_{i}(\bold{r}) = \phi^{\bold{k}}_{i}(\bold{S^{-1}r}) = \sum_{j} d_{ji}(\bold{S},\bold{k}) \phi^{\bold{kS^{-1}}}_{j}(\bold{r}).  
\label{eq:bloch_sym}
\end{align}

The derivation presented in this work and implemented in our code is for a general single-particle basis and suitable for different approaches to the electronic structure problem, including plane-waves, atom-centered localized orbitals, KS states in any suitable basis, among others. In the case of the symmetry-adapted ISDF implementation, we do require that Eq.~\ref{eq:bloch_sym} is satisfied in order to have an exact representation of an otherwise equivalent calculation without the use of symmetry. In practice, when the single-particle basis comes from a truncated set of MF solutions, all degenerate sets of orbitals must be included in order to not violate Eq.~\ref{eq:bloch_sym}. While discarding states in degenerate blocks will lead to errors, in practice these are small and are furthermore reduced as the number of virtual states is increased, since this will typically only happen for a handful of states at the top of the set of virtual bands. For exact eigenstates of a symmetric Hamiltonian, $\bold{d}(\bold{S},\bold{k})$ will form an irreducible representation of the group \textit{G}, but in order to accommodate a more general single-particle basis we do not assume that it is irreducible. Instead we calculate the full $\bold{d}(\bold{S},\bold{k})$ in the general case but store it as a sparse tensor to benefit from potential memory and computational savings.

It is easy to show that in the basis of single-particle orbitals satisfying Eq.~\ref{eq:bloch_sym}, totally symmetric one-body operators transform according to $O^{\bold{k}}_{ij} = \sum_{ab}d^{*}_{ai}(\bold{S},\bold{k}) O^{\bold{kS^{-1}}}_{ab} d_{bj}(\bold{S},\bold{k})$,  including one-electron Hamiltonians, self energies, etc. In our implementation, we use this relation to reconstruct operators in the full BZ from those within the irreducible sector. We also exploit Eq.~\ref{eq:bloch_sym} to avoid storing the basis states in the full BZ, explicitly storing them only in the irreducible BZ (IBZ) and taking $d(\bold{S},\bold{k})_{ij} = \delta_{ij}$ for the specific symmetry operation chosen to represent a given $\mathbf{k}$-point outside the full BZ and the corresponding point inside the IBZ. Notice that, for any other symmetry and $\mathbf{k}$-point combination, $\bold{d}(\bold{S'},\bold{k})$ would need to be evaluated and stored.  

\section{Symmetry-adapted interpolative separable density fitting\label{sec:isdf}}
Given a set of single-particle basis functions $\{\phi^{\bold{k}}_{i}(\bold{r})\}$, the Schrödinger equation can be reformulated in the second quantization language. In this formulation, the electron-electron Coulomb kernel is expanded using the product basis $\phi^{\textbf{k}_{1}}_{i}(\textbf{r})\phi^{\textbf{k}_{2}*}_{j}(\textbf{r})$, resulting in the ERI tensor: 
\begin{align}
V^{\textbf{k}_{1}\textbf{k}_{2}\textbf{k}_{3}\textbf{k}_{4}}_{\ i\ j\ \ k\ l} = 
\int d\textbf{r} \int d\textbf{r}' \phi^{\textbf{k}_{i}*}_{1}(\textbf{r})\phi^{\textbf{k}_{2}}_{j}(\textbf{r})\frac{1}{|\textbf{r}-\textbf{r}'|}\phi^{\textbf{k}_{3}*}_{k}(\textbf{r}')\phi^{\textbf{k}_{4}}_{l}(\textbf{r}').   
\label{eq:eri}
\end{align}
Here, the momentum transferred between the product basis (also known as the pair densities) satisfies the conservation of momentum $\bold{k}_{1} - \bold{k}_{2} + \bold{G} = \bold{k}_{3} - \bold{k}_{4}$, where $\bold{G}$ is a reciprocal lattice vector. Although the number of orbitals ($N_{\mathrm{orb}}$) is typically much smaller than the dimension of real-space discretization, the integrals in Eq.~\ref{eq:eri} destroy the separability of the product basis in the orbital and momentum indices, resulting in a fourth-order tensor. Notice that in general, this tensor is strongly rank deficient, reflecting the over-completeness of the product basis for two-electron operators. Many-body methods that directly operate on the uncompressed form of the ERIs lead to unfavorable computational complexity concerning $N_{k}$ and $N_{\mathrm{orb}}$.

Following the $\bold{q}$-resolved ISDF procedure introduced in Ref.~\onlinecite{THC-RPA_CNY2023}, we now derive how space-group symmetries of crystals can be incorporated into ISDF while maintaining the full separability of the pair densities in the resulting decomposed ERIs. 

Considering a symmetry operation $\bold{S}$ that maps a transferred momentum $\bold{q}$ to $\bar{\bold{q}} = \bold{q}\bold{S}$ inside the IBZ (for simplicity, we will use the overline symbol to denote a crystal momentum is inside the IBZ), the pair densities $\phi^{\bold{k-q}*}_{j}(\bold{r})\phi^{\bold{k}}_{i}(\bold{r})$ can be related to the symmetry-related ones $\phi^{\bold{kS-\bar{q}}*}_{a}(\bold{S}^{-1}\bold{r})\phi^{\bold{kS}}_{b}(\bold{S}^{-1}\bold{r})$  through basis set transformations and then expanded through the $\bold{q}$-resolved ISDF as: 
\begin{subequations}
\begin{align}
&\phi^{\bold{k-q}*}_{j}(\bold{r})\phi^{\bold{k}}_{i}(\bold{r}) \label{eq:isdf_sym_a}\\
&= \sum_{ab}d^{*}_{aj}(\bold{S}^{-1},\bold{k-q})d_{bi}(\bold{S}^{-1},\bold{k})\label{eq:isdf_sym_b}\\
&\ \ \ \ \ \ \ \ \ \ \ \ \ \ \ \ \ \ \ \ \ \ \times \phi^{\bold{kS-\bar{q}}*}_{a}(\bold{S}^{-1}\bold{r})\phi^{\bold{kS}}_{b}(\bold{S}^{-1}\bold{r}) \nonumber\\
&= \sum_{ab}d^{*}_{aj}(\bold{S}^{-1},\bold{k-q})d_{bi}(\bold{S}^{-1},\bold{k})\label{eq:isdf_sym_c}\\
&\ \ \ \ \ \ \ \ \ \ \ \ \ \times \sum_{\mu}\phi^{\bold{kS-\bar{q}}*}_{a}(\bold{r}_{\mu})\phi^{\bold{kS}}_{b}(\bold{r}_{\mu})\zeta^{\bold{\bar{q}}}_{\mu}(\bold{S}^{-1}\bold{r}) \nonumber \\
&= \sum_{\mu}\phi^{\bold{k-q}*}_{j}(\bold{S}\bold{r}_{\mu})\phi^{\bold{k}}_{i}(\bold{S}\bold{r}_{\mu})\zeta^{\bold{\bar{q}}}_{\mu}(\bold{S}^{-1}\bold{r}). \label{eq:isdf_sym_d}
\end{align}
\label{eq:isdf_sym}
\end{subequations}
Eq.~\ref{eq:isdf_sym_c} assumes the existence of a $\bold{q}$-resolved ISDF decomposition at $\bold{\bar{q}}$:
\begin{align}
\phi^{\bold{k-\bar{q}}*}_{a}(\bold{r})\phi^{\bold{k}}_{b}(\bold{r}) = \sum_{\mu} \phi^{\bold{k-\bar{q}}*}_{a}(\bold{r}_{\mu})\phi^{\bold{k}}_{b}(\bold{r}_{\mu})\zeta^{\bold{\bar{q}}}_{\mu}(\bold{r}), 
\label{eq:isdf}
\end{align}
where $\{\bold{r}_{\mu}\}$ are the ISDF interpolating points and $\{\zeta^{\bar{\bold{q}}}_{\mu}\}$ are the ISDF auxiliary functions. 
Eqs.~\ref{eq:isdf_sym} suggest that considering the ISDF interpolating vectors with the transferred momenta $\bold{\bar{q}}$ inside the IBZ is enough to expand all the pair densities. 
Nevertheless, the price is either the additional matrix multiplications with $\bold{d}(\bold{S}^{-1},\bold{k})$ as shown in Eq.~\ref{eq:isdf_sym_c} or the symmetry-dependent interpolated points $\{\bold{S}\bold{r}_{\mu}\}$ as shown in Eq.~\ref{eq:isdf_sym_d}. 

The symmetry-adapted ISDF can be summarized as follows: 
Given a set of interpolating points $\{\bold{r}_{\mu}\}$, we solve the least-squares problem of Eq.~\ref{eq:isdf} to obtain the ISDF auxiliary basis $\zeta^{\bold{\bar{q}}}_{\mu}(\bold{r})$ only for the irreducible $\bold{q}$-points. 
The details of solving Eq.\ref{eq:isdf} can be found in Sec. 3.1 in Ref.~\onlinecite{THC-RPA_CNY2023}. 
Next, a fully separable representation of ERIs for $\bold{q}$-points over the entire BZ can be constructed as:
\begin{subequations}
\begin{align}
V&^{\textbf{k}_{1}\textbf{k}_{2}\textbf{k}_{3}\textbf{k}_{4}}_{\ i\ j\ \ k\ l} = \sum_{\mu\nu}\phi^{\bold{k}_{1}*}_{i}(\bold{S}\bold{r}_{\mu})\phi^{\bold{k}_{2}}_{j}(\bold{S}\bold{r}_{\mu})\label{eq:thc_sym_a}\\
&\ \ \ \ \ \ \ \ \ \ \ \ \ \ \ \ \ \ \ \ \ \ \ \ \times V^{\bold{qS}}_{\mu\nu}\phi^{\bold{k}_{3}*}_{k}(\bold{S}\bold{r}_{\nu})\phi^{\bold{k}_{4}}_{l}(\bold{S}\bold{r}_{\nu}) \nonumber\\
&=\sum_{\mu\nu}X^{\bold{k}_{1}*}_{\mu i}(\bold{S})X^{\bold{k}_{2}}_{\mu j}(\bold{S}) V^{\bold{\bar{q}}}_{\mu\nu}X^{\bold{k}_{3}}_{\nu k}(\bold{S})X^{\bold{k}_{4}}_{\nu l}(\bold{S}), \label{eq:thc_sym_b}
\end{align}
\label{eq:thc_sym}
\end{subequations}
where $\bold{q} = \bold{k}_{1} - \bold{k}_{2} + \bold{G}$, $\bold{S}$ is the rotation matrix that maps $\bold{q}$ to $\bold{\bar{q}} = \bold{qS}$ inside the IBZ, and:
\begin{align}
&X^{\bold{k}_{1}}_{\mu i}(\bold{S}) = \phi^{\bold{k}_{1}}_{u}(\bold{S}\bold{r}_{\mu}), \label{eq:ISDF_X}\\
&V^{\bold{\bar{q}}}_{\mu\nu} = \int d\textbf{r} \int d\textbf{r}' \zeta^{\bar{\textbf{q}}*}_{\mu}(\textbf{r})\frac{1}{|\textbf{r}-\textbf{r}'|}\zeta^{\bar{\textbf{q}}}_{\nu}(\textbf{r}'). \label{eq:ISDF_Vq} 
\end{align} 
The absence of the $\bold{q}$ dependence of ${\bold{r}_{\mu}}$ is crucial for the realization of the THC-like representation of ERIs in Eqs.~\ref{eq:thc_sym_b}. 
Unfortunately, the dependence of $\bold{S}$ in the collocation matrix $\bold{X}^{\bold{k}}$ is inevitable due to the symmetry-dependent interpolating points. 
In practice, we choose $\{\bold{r}_{\mu}\}$ to be the interpolating points obtained at $\bold{q}=\bold{0}$ which turns out to provide good accuracy as analyzed in Ref.~\onlinecite{THC-RPA_CNY2023}.

Since the rank of the pair densities grows linearly with respect to $N_{\mathrm{orb}}$, the size of the ISDF interpolating points ($N_{\mu}$) is expected to be $O(N_{\mathrm{orb}})$ as well. 
In practice, it has been demonstrated that $N_{\mu} = \alpha N_{\mathrm{orb}}$ with $\alpha\sim8-12$ is sufficiently effective in achieving chemical accuracy~\cite{THC-RPA_CNY2023}.
In the complete basis set limit, where $N_{\mathrm{orb}}$ equals the size of the real-space discretization $N_{r}$, the maximum number of interpolating points should also be $N_{r}$, the maximum rank of the pair density $\phi^{\bold{k-q}*}_{i}(\bold{r})\phi^{\bold{k}}_{j}(\bold{r})$. Consequently, as $N_{\mathrm{orb}}$ increases, one would expect $\alpha$ to become smaller, eventually reducing to one.

The applicability of ISDF should not be limited to the factorization of ERIs. 
Instead, it provides a general prescription for constructing a two-particle basis for generic single-particle orbitals. 
As such, it is capable of compressing any two-particle operator while retaining separability in the orbital and momentum indices. 
Another application of ISDF lies in the context of quantum embedding, where basis transformations between the full Hilbert space and the downfolded subspace are executed frequently. 
As discussed in Sec.~\ref{subsec:local_pb}, the ISDF auxiliary basis can act as a proxy between the two Hilbert spaces, accelerating the construction of low-energy Hamiltonians. 

\section{MBPT within the symmetry-adapted ISDF formalism \label{sec:isdf_mbpt}} 

\subsection{ISDF-\texorpdfstring{$GW$}{GW} \label{subsec:isdf_gw}}
In this subsection, we derive a low-scaling algorithm for $GW$ in a canonical basis using the symmetry-adapted ISDF technique. 
The resulting algorithm shares the same complexity as the space-time formalism: cubic scaling in the system sizes and linear scaling in the number of $\textbf{k}$-points. 
Furthermore, the prefactors are much smaller in ISDF-$GW$, depending on the size of the ISDF auxiliary basis. 

We begin with the $GW$ self-energy expressed in a canonical basis:
\begin{align}
\Sigma^{\bar{\bold{k}}}_{ij}(\tau) = -\frac{1}{N_{q}}\sum_{\bold{q}\in\text{BZ}}\sum_{ab}G^{\bold{\bar{k}-q}}_{ab}(\tau)W^{\bar{\bold{k}},\bold{\bar{k}-q},\bold{\bar{k}-q},\bar{\bold{k}}}_{i,a,b,j}(\tau)
\label{eq:gw_sigma_canonical}
\end{align}
where $N_{q}$ is the size of the $\bold{q}$-point discretization, and the screened interaction $W$ is expanded using the product basis: 
\begin{align}
W&^{\textbf{k}_{1}\textbf{k}_{2}\textbf{k}_{3}\textbf{k}_{4}}_{\ i\ j\ \ k\ l}(\tau) = \\
&\int d\textbf{r} \int d\textbf{r}' \phi^{\textbf{k}_{1}*}_{i}(\textbf{r})\phi^{\textbf{k}_{2}}_{j}(\textbf{r})W(\bold{r},\bold{r}';\tau)\phi^{\textbf{k}_{3}*}_{k}(\textbf{r}')\phi^{\textbf{k}_{4}}_{l}(\textbf{r}'). \nonumber
\end{align}

Applying symmetry-related ISDF to the product basis leads to a THC-like factorization of $W$:
\begin{align}
W^{\bold{k}_{1}\bold{k}_{2}\bold{k}_{3}\bold{k}_{4}}_{\ i\ j\ \ k\ l}(\tau) &= \sum_{\mu\nu} \phi^{\bold{k}_{1}*}_{i}(\bold{S}\bold{r}_{\mu})\phi^{\bold{k}_{2}}_{j}(\bold{S}\bold{r}_{\mu})\label{eq:bloch_W_thc}\\
& \times W^{\bold{\bar{q}}}_{\mu\nu}(\tau)\phi^{\bold{k}_{3}*}_{k}(\bold{S}\bold{r}_{\nu})\phi^{\bold{k}_{4}}_{l}(\bold{S}\bold{r}_{\nu}), \nonumber
\end{align}
where $\bold{\bar{q}}$ and $\bold{S}$ are defined in a similar manner as in Eqs.~\ref{eq:thc_sym}, and the matrix elements of $W$ in the ISDF auxiliary basis are defined as:
 \begin{align}
W^{\bold{\bar{q}}}_{\mu\nu}(\tau) = \int d\bold{r} \int d\bold{r}' \zeta^{\bold{\bar{q}}*}_{\mu}(\bold{r})W(\bold{r}, \bold{r}'; \tau) \zeta^{\bold{\bar{q}}}_{\nu}(\bold{r}').
\label{eq:W_munu}
\end{align} 

Inserting Eq.~\ref{eq:bloch_W_thc} into Eq.~\ref{eq:gw_sigma_canonical}, we arrive at our self-energy formula for ISDF-$GW$: 
\begin{widetext}
\begin{subequations}
\begin{align}
\Sigma^{\bar{\bold{k}}}_{ij}&(\tau) = -\frac{1}{N_{q}}\sum_{\bold{q}\in\text{BZ}}\sum_{ab}G^{\bold{\bar{k}-q}}_{ab}(\tau)W^{\bar{\bold{k}},\bold{\bar{k}-q},\bold{\bar{k}-q},\bar{\bold{k}}}_{i,a,b,j}(\tau) \label{eq:GW_sym_a}\\ 
&= -\frac{1}{N_{q}}\sum_{\bold{S}}\sum_{\bar{\bold{q}}\in\text{IBZ}}^{N_{\bar{\bold{q}}/\bold{S}}}\sum_{ab}G^{\bold{\bar{k}-\bar{q}S}^{-1}}_{ab}(\tau)\sum_{\mu\nu}\phi^{\bar{\bold{k}}*}_{i}(\bold{S}\bold{r}_{\mu})\phi^{\bold{\bar{k}-\bar{q}S}^{-1}}_{a}(\bold{S}\bold{r}_{\mu})W^{\bar{\bold{q}}}_{\mu\nu}(\tau)\phi^{\bold{\bar{k}-\bar{q}S}^{-1}*}_{b}(\bold{S}\bold{r}_{\nu})\phi^{\bar{\bold{k}}}_{j}(\bold{S}\bold{r}_{\nu}) \label{eq:GW_sym_b} \\
&= -\frac{1}{N_{q}}\sum_{\bold{S}} \sum_{\mu\nu}\phi^{\bar{\bold{k}}*}_{i}(\bold{S}\bold{r}_{\mu}) \phi^{\bar{\bold{k}}}_{j}(\bold{S}\bold{r}_{\nu}) \sum_{\bar{\bold{q}}\in\text{IBZ}}^{N_{\bar{\bold{q}}/\bold{S}}} G^{\bold{\bar{k}-\bar{q}S}^{-1}}(\bold{S}\bold{r}_{\mu}, \bold{S}\bold{r}_{\nu}; \tau)W^{\bar{\bold{q}}}_{\mu\nu}(\tau) \label{eq:GW_sym_c}\\
&= -\frac{1}{N_{q}}\sum_{\bold{S}} \sum_{\mu\nu}\phi^{\bar{\bold{k}}*}_{i}(\bold{S}\bold{r}_{\mu}) \phi^{\bar{\bold{k}}}_{j}(\bold{S}\bold{r}_{\nu}) \sum_{\bar{\bold{q}}\in\text{IBZ}}^{N_{\bar{\bold{q}}/\bold{S}}} G^{\bold{\bar{k}S}-\bold{\bar{q}}}(\bold{r}_{\mu}, \bold{r}_{\nu}; \tau)W^{\bar{\bold{q}}}_{\mu\nu}(\tau) \label{eq:GW_sym_d} \\
&=\sum_{\bold{S}}\sum_{\mu\nu}\phi^{\bar{\bold{k}}*}_{i}(\bold{S}\bold{r}_{\mu}) \phi^{\bar{\bold{k}}}_{j}(\bold{S}\bold{r}_{\nu})\Sigma^{\bar{\bold{k}}}_{\mu\nu}(\bold{S};\tau), \label{eq:GW_sym_e} 
\end{align}
\label{eq:GW_sym}
\end{subequations}
\end{widetext}
where we have changed the variables from $\bold{q}$ to $\bar{\bold{q}}\bold{S}$, and the summation over the entire $\bold{q}$ discretization is split into:
\begin{align}
\sum_{\bold{q}\in\mathrm{BZ}} \rightarrow \sum_{\bold{S}}\sum_{\bar{\bold{q}}\in\mathrm{IBZ}}^{N_{\bar{\bold{q}}/\bold{S}}}, 
\end{align}
where $N_{\bar{\bold{q}}/\bold{S}}$ denotes the number of irreducible $\bold{q}$-points satisfying $\bar{\bold{q}} = \bold{qS} \in$ IBZ, and it can vary for different $\bold{S}$ ($1 \le N_{\bar{q}/\bold{S}} \le N_{\bar{q}}$). 
Since, in principle, a $\bold{q}$-point can be mapped to different points inside the IBZ through different symmetry rotations, it is necessary to manually exclude those double-counted $\bold{q}$-points. 

In Eqs.~\ref{eq:GW_sym_c} and~\ref{eq:GW_sym_d}, $G^{\bar{\bold{k}}-\bar{\bold{q}}\bold{S}^{-1}}(\bold{S}\bold{r}_{\mu}, \bold{S}\bold{r}_{\nu}; \tau)$ is the Green's function evaluated on the rotated ISDF interpolating points. This can be expressed as:
\begin{subequations}
\begin{align}
G^{\bar{\bold{k}}}(\bold{S}\bold{r}_{\mu}, \bold{S}\bold{r}_{\nu};\tau) &= \sum_{ab}\phi^{\bar{\bold{k}}}_{a}(\bold{S}\bold{r}_{\mu})G^{\bar{\bold{k}}}_{ab}(\tau)\phi^{\bar{\bold{k}}*}_{b}(\bold{S}\bold{r}_{\nu}) \label{eq:G_thc_a}\\
&= G^{\bar{\bold{k}}\bold{S}}(\bold{r}_{\mu}, \bold{r}_{\nu}; \tau), 
\end{align}
\label{eq:G_thc}
\end{subequations}
where the second equality is based on Eq~\ref{eq:bloch_sym}.

In Eq.~\ref{eq:GW_sym_e}, $\Sigma^{\bar{\bold{k}}}_{\mu\nu}(\bold{S}; \tau)$ is defined as the self-energy contribution at a given symmetry operation $\bold{S}$:
\begin{align}
\Sigma^{\bar{\bold{k}}}_{\mu\nu}(\bold{S}; \tau) = -\frac{1}{N_{q}}\sum_{\bar{\bold{q}}\in\text{IBZ}}^{N_{\bar{\bold{q}}/\bold{S}}} G^{\bar{\bold{k}}\bold{S}-\bold{\bar{q}}}(\bold{r}_{\mu}, \bold{r}_{\nu}; \tau)W^{\bar{\bold{q}}}_{\mu\nu}(\tau). 
\label{eq:isdf_gw_sigma_S}
\end{align}
In the absence of space-group symmetry, the $GW$ self-energy in the ISDF auxiliary basis simplifies to:
\begin{align}
\Sigma^{\bar{\bold{k}}}_{\mu\nu}(\tau) = -\frac{1}{N_{q}}\sum_{\bold{q}\in\mathrm{BZ}}G^{\bar{\bold{k}}-\bold{q}}(\bold{r}_{\mu}, \bold{r}_{\nu}; \tau) W^{\bold{q}}_{\mu\nu}(\tau), 
\end{align}
where the momentum convolution can be avoided in real space:
\begin{align}
\Sigma^{\bold{R}}_{\mu\nu}(\tau) = -G^{\bold{R}}(\bold{r}_{\mu}, \bold{r}_{\nu}; \tau)W^{\bold{R}}_{\mu\nu}(\tau). 
\label{eq:isdf_gw_sigma_nosym}
\end{align}

It is evident that Eq.~\ref{eq:isdf_gw_sigma_nosym} shares the same structure as in the space-time formalism (Eq.~\ref{eq:gw_spacetime_e}). 
However, ISDF-$GW$ requires information only on $\{\bold{r}_{\mu}\}$, whose dimension is typically orders of magnitude smaller than a uniform real-space grid employed in the space-time formalism. 

The remaining piece is to compute $W^{\bar{\bold{q}}}_{\mu\nu}(\tau)$. 
By projecting Eq.~\ref{eq:hedin_dysonW} onto the ISDF auxiliary basis, we obtain the Dyson equation for the screened interaction in a compact two-particle basis:
\begin{align}
\bold{W}^{\bold{\bar{q}}}(i\Omega_{n}) = \tilde{\bold{W}}^{\bar{\bold{q}}}(i\Omega_{n}) + \bold{V}^{\bold{\bar{q}}} = [\bold{I} - \bold{V}^{\bold{\bar{q}}}\boldsymbol{\Pi}^{\bold{\bar{q}}}(i\Omega_{n})]^{-1}\bold{V}^{\bold{\bar{q}}}, \label{eq:ISDF_W_dyson}
\end{align}
where $i\Omega_{n}$ denotes bosonic Matsubara frequencies, $\bold{I}$ represents the identity matrix, $\bold{V}^{\bar{\bold{q}}}$ is the bare Coulomb matrix defined in Eq.~\ref{eq:ISDF_Vq}, $\tilde{\bold{W}}^{\bar{\bold{q}}}(i\Omega_{n})$ is the dynamic part of the screened interaction, and $\boldsymbol{\Pi}^{\bold{\bar{q}}}(i\Omega_{n})$ is the irreducible polarizability evaluated at $\{\bold{r}_{\mu}\}$:
\begin{subequations}
\begin{align}
\Pi^{\bold{\bar{q}}}_{\mu\nu}&(\tau) = \Pi^{\bold{\bar{q}}}(\bold{r}_{\mu}, \bold{r}_{\nu}) \label{eq:ISDF_Pi_a}\\
&= \frac{1}{N_{k}}\sum_{\bold{k}\in\mathrm{BZ}}G^{\bold{k}}(\bold{r}_{\mu}, \bold{r}_{\nu}; \tau)G^{\bold{k-\bar{q}}}(\bold{r}_{\nu}, \bold{r}_{\mu}; -\tau) \label{eq:ISDF_Pi_b}\\
&=G^{\bold{R}}(\bold{r}_{\mu}, \bold{r}_{\nu}; \tau)G^{-\bold{R}}(\bold{r}_{\nu}, \bold{r}_{\mu}; -\tau). \label{eq:ISDF_Pi_c}
\end{align}
\label{eq:ISDF_Pi}
\end{subequations}

In practical computation, the Fourier transform of $\bold{W}^{\bar{\bold{q}}}(i\Omega_{n})$ to the imaginary-time axis yields a delta function at $\tau=0^{-}$ due to the static contribution from the bare interaction, as illustrated in Eq.~\ref{eq:ISDF_W_dyson}. 
Consequently, it is convenient to decompose the $GW$ self-energy in Eqs.~\ref{eq:GW_sym} into static and dynamic components: 
\begin{align}
\Sigma^{\bar{\bold{k}}}_{ij}(\tau) = K^{\bar{\bold{k}}}_{ij} + \tilde{\Sigma}^{\bold{\bar{k}}}_{ij}(\tau). 
\end{align}
The static part of the $GW$ self-energy ($K^{\bar{\bold{k}}}$) is equivalent to the exchange potential from the HF approximation. 
Finally, Eqs.~\ref{eq:GW_sym},~\ref{eq:G_thc},~\ref{eq:ISDF_W_dyson}, and~\ref{eq:ISDF_Pi} collectively constitute the complete ISDF-$GW$ equations.

\subsubsection{Complexity Analysis\label{subsubsec:compexity}}
We analyze the computational complexity by breaking down the ISDF-$GW$ equations into three distinct parts: the evaluation of $\Pi$, the evaluation of $W$, and the evaluation of $\Sigma$. 
Depending on the dimensions of the target system, the computation will be dominated by different parts. 

The evaluation of $\Pi$ includes Eqs.~\ref{eq:ISDF_Pi} and the computation of the Green's function $G^{\bold{k}}(\bold{r}_{\mu},\bold{r}_{\nu}; \tau)$ at the interpolating points. 
Regardless of whether space-group symmetries are considered, $G^{\bold{k}}(\bold{r}_{\mu},\bold{r}_{\nu}; \tau)$ needs to be computed for all $\bold{k}$-points sampled in the first BZ. 
This operation scales as $O(N_{\tau}N_{k}N_{\mu}^{2}N_{\mathrm{orb}})$, where $N_{\tau}$ corresponds to the size of the $\tau$ discretization. 
Once $G^{\bold{k}}(\bold{r}_{\mu},\bold{r}_{\nu}; \tau)$ is computed, Eqs.~\ref{eq:ISDF_Pi} can be evaluated at a cost of $O(N_{\tau}N_{k}\ln{N_{k}}N_{\mu}^{2})$ using the FFT to circumvent the convolution in momentum space.

Solving the Dyson equation for $W$ in the ISDF auxiliary basis scales as $O(N_{\Omega}N_{\bar{q}}N_{\mu}^{3})$, where $N_{\Omega}$ is the number of bosonic Matsubara frequencies, and $N_{\bar{q}}$ is the number of irreducible $\bold{q}$-points. 

The computation of $\Sigma$ (Eqs.~\ref{eq:GW_sym}) is primarily influenced by Eqs.~\ref{eq:G_thc},~\ref{eq:isdf_gw_sigma_S}, and~\ref{eq:GW_sym_e}. 
Similar to the evaluation of $\Pi$, assessing $G^{\bold{k}-\bar{\bold{q}}\bold{S}^{-1}}(\bold{r}_{\mu},\bold{r}_{\nu}; \tau)$ scales as $O(N_{\tau}N_{k}N_{\mu}^{2}N_{\mathrm{orb}})$. This is because, in principle, $\bold{k}-\bar{\bold{q}}\bold{S}^{-1}$ can extend beyond the IBZ for various combinations of $\bar{\bold{k}}$, $\bar{\bold{q}}$, and $\bold{S}$.
Subsequently, Eqs.~\ref{eq:GW_sym_e} and~\ref{eq:isdf_gw_sigma_S} can be computed at the expense of $O(N_{\tau}N_{\bar{k}}N_{\bold{S}}N_{\mu}^{2}N_{\mathrm{orb}})$ and $O(N_{\tau}N_{\bar{k}}N_{\bold{S}}N_{\bar{q}/\bold{S}}N_{\mu}^{2})$. 
Here, $N_{\bar{k}}$ represents the number of irreducible $\bold{k}$-points, $N_{\bold{S}}$ is the number of considered symmetry operations, and $N_{\bar{q}/\bold{S}}$ is the number of irreducible $\bold{q}$-points per symmetry operation $\bold{S}$. 
In the absence of space-group symmetry ($N_{\bold{S}}=1$ and $N_{\bar{q}/\bold{S}}=N_{q}$), Eq.~\ref{eq:isdf_gw_sigma_S} can be solved in real space, as demonstrated in Eq.~\ref{eq:isdf_gw_sigma_nosym}, at a cost of $O(N_{\tau}N_{k}\ln{N_{k}}N_{\mu}^{2})$.

Assuming $N_{\mathrm{orb}} > N_{k}$, and $N_{\tau}=N_{\Omega}$, the dominant step in the ISDF-$GW$ equations will depends on the number of symmetry operations present in the target system. 
In the absence of space-group symmetry, the computation of the screened interaction becomes the most computationally demanding, due to its cubic scaling with respect to $N_{\mu}$. 
As the number of symmetry operations increases, Eq.~\ref{eq:ISDF_W_dyson} experiences a significant speedup, while the costs associated with the evaluations of the irreducible polarizability and the self-energy remain relatively constant. This is attributed to the necessity of $G^{\bold{k}}(\bold{r}_{\mu}, \bold{r}_{\nu}; \tau)$ over the full BZ and the explicit dependence on $N_{\bold{S}}$ in Eqs.~\ref{eq:GW_sym_e} and~\ref{eq:isdf_gw_sigma_S}.
Regardless, our analysis indicates that symmetry-adapted ISDF-$GW$ formally scales cubically with respect to the system sizes and linearly with respect to the number of $\bold{k}$-points.

From the perspective of memory usage, the dominant components are the bosonic quantities expanded in the ISDF auxiliary basis, which include the screened interactions $W^{\bold{\bar{q}}}_{\mu\nu}(\tau)$ and the irreducible polarizability $\Pi^{\bar{\bold{q}}}_{\mu\nu}(\tau)$. It's noteworthy that storing the entire $G^{\bold{k}}(\bold{r}_{\mu}, \bold{r}_{\nu}; \tau)$ tensor can be avoided by splitting the calculations sequentially along the $\tau$ index.
This is possible since $G^{\bold{k}}(\bold{r}_{\mu}, \bold{r}_{\nu}; \tau)$ is an intermediate quantity in Eqs.~\ref{eq:GW_sym} and~\ref{eq:ISDF_Pi}. 
Since the bosonic quantities are only required at the irreducible $\bold{q}$-points, the memory load benefits from the inclusion of space-group symmetries. 
Overall, the memory requirement for ISDF-$GW$ scales as $O(N_{\tau}N_{\bar{q}}N_{\mu}^{2})$. 

\begin{figure*}[tbh!]
\begin{center}
\includegraphics[width=0.65\textwidth]{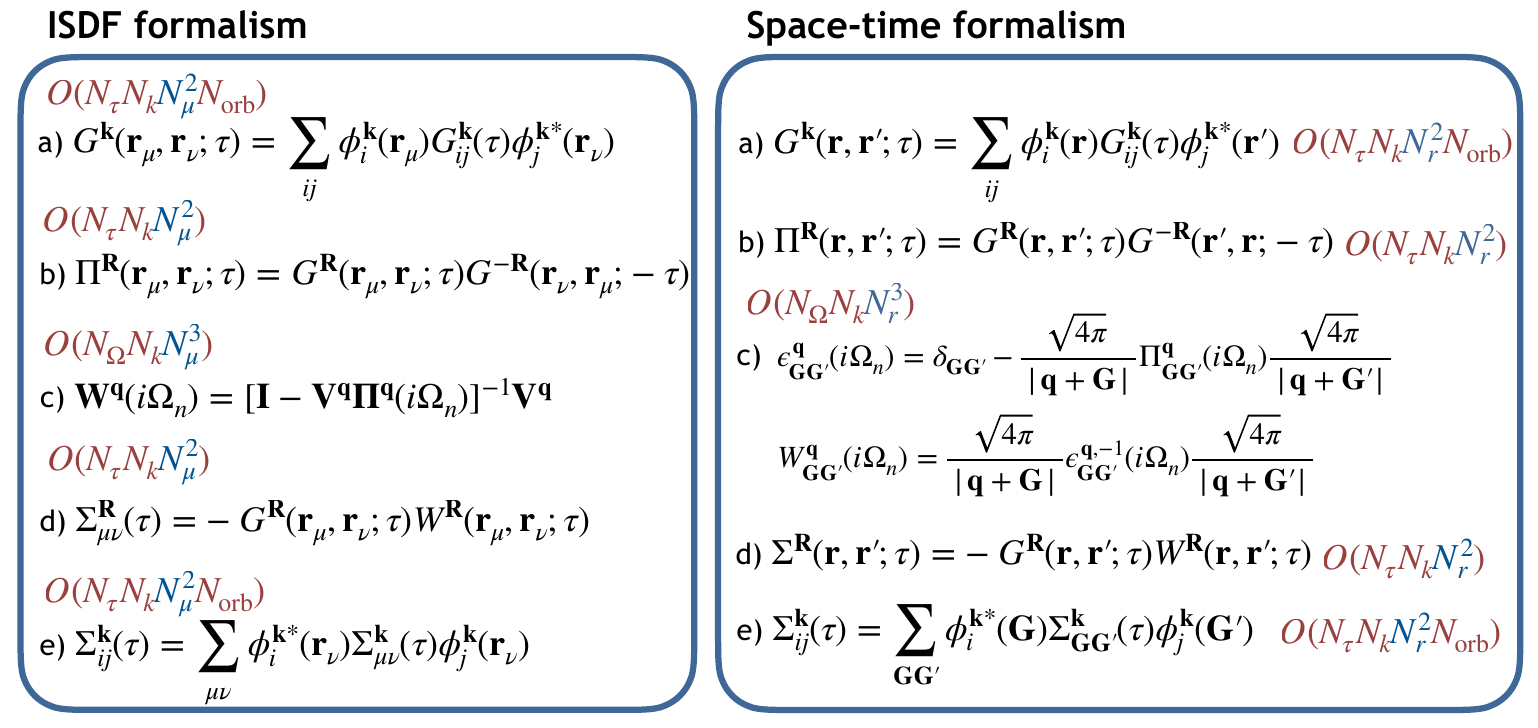} 
\caption{Comparisons of the $GW$ equations between the ISDF and the space-time formalism}
\label{fig:gw_eqns_compare}
\end{center}
\end{figure*}

Finally, we would like to draw a side-by-side comparison between ISDF-$GW$ and the space-time formalism. 
For simplicity, we consider the limit of no space-group symmetry employed. 
Illustrated in Fig.~\ref{fig:gw_eqns_compare}, the steps involved in computing $\Pi$ and $\Sigma$ in both formalisms share a similar structure where the evaluations are both executed in space-time coordinates.
However, the dimensions in the two formalisms vary significantly: the size of $\{\bold{r}_{\mu}\}$ is typically orders of magnitude smaller than a uniform real-space discretization. 
This discrepancy results in a considerably smaller prefactor in ISDF-$GW$ computations.

Since $N_{\mu}$ is proportional to $N_{\mathrm{orb}}$, the ratio of $N_{\mu}$ to the size of the real-space grid employed in FFT ($N_{r}$) gradually approaches one as $N_{\mathrm{orb}}$ increases. 
Consequently, the speedup against the space-time formalism diminishes. 
ISDF-$GW$ eventually becomes equivalent to the space-time formalism when $N_{\mu} = N_{r}$. 
However, in cases where $N_{\mathrm{orb}} < N_{r}$, ISDF automatically determines a non-uniform set of real-space grid points for the given single-particle basis set, achieving a specified level of accuracy. 
As long as $N_{\mathrm{orb}} \ll N_{r}$, we anticipate a substantial performance improvement of ISDF-$GW$ over the space-time formalism. This efficiency is particularly notable for calculations with hard pseudopotentials or low-lying electrons. Significant performance improvements can also be obtained when combined with techniques designed to accelerate convergence with respect to the number of virtual orbitals~\cite{GW_virtuals_Bruneval2008,GW_virutals_Gao2016}, since this will also lead to smaller values of $N_{\mathrm{orb}}$. This will be the subject of future work.

\subsection{Local product basis \label{subsec:local_pb}}
In the context of quantum embedding, a material-specific low-energy Hamiltonian is formulated for a strongly correlated subspace $\mathcal{C}$. 
Typically, $\mathcal{C}$ is defined by a set of local basis functions $\{w_{\alpha}(\bold{r})\}$ which, in practice, are constructed as linear combinations of the primary Bloch basis $\{\phi^{\bold{k}}_{i}(\bold{r})\}$:
\begin{align}
w_{\alpha}(\bold{r}) = \frac{1}{N_{k}}\sum_{\bold{k}}\sum_{i}C^{\bold{k}*}_{\alpha i}\phi^{\bold{k}}_{i}(\bold{r})e^{-i\dprod{k}{R}}\big|_{\bold{R=0}}. 
\end{align}
Here, $C^{\bold{k}}_{\alpha i} = \langle w^{\bold{k}}_{\alpha}| \phi^{\bold{k}}_{i}\rangle$ denotes the orbital transformation, and $\bold{R}$ represents a translational vector of the system.

Similarly, bosonic quantities are expanded using the local product basis:  
\begin{align}
\Psi_{\alpha\beta}(\bold{r}) = w^{*}_{\alpha}(\bold{r})w_{\beta}(\bold{r}) = \frac{1}{N_{q}}\sum_{\bold{q}}\Psi^{\bold{q}}_{\alpha\beta}(\bold{r})e^{-i\dprod{q}{R}}\big|_{\bold{R=0}}.  
\label{eq:local_pb}
\end{align}
The $\bold{q}$-resolved product basis $\Psi^{\bold{q}}_{\alpha\beta}(\bold{r})$ can be expanded using the ISDF auxiliary basis:
\begin{subequations}
\begin{align}
&\Psi^{\bold{q}}_{\alpha\beta}(\bold{r}) = \frac{1}{N_{k}}\sum_{\bold{k}}\sum_{ij}C^{\bold{k-q}}_{\alpha i}C^{\bold{k}*}_{\beta j}\phi^{\bold{k-q}*}_{i}(\bold{r})\phi^{\bold{k}}_{j}(\bold{r}) \label{eq:lpb_a}\\ 
&= \frac{1}{N_{k}}\sum_{\bold{k}}\sum_{ij\mu}C^{\bold{k-q}}_{\alpha i}C^{\bold{k}*}_{\beta j}\phi^{\bold{k-q}*}_{i}(\bold{S}\bold{r}_{\mu})\phi^{\bold{k}}_{j}(\bold{S}\bold{r}_{\mu})\zeta^{\bold{\bar{q}}}_{\mu}(\bold{S}^{-1}\bold{r}) \label{eq:lpb_q_b}\\ 
&= \sum_{\mu} \Psi^{\bold{q}}_{\alpha\beta}(\bold{S}\bold{r}_{\mu})\zeta^{\bold{\bar{q}}}_{\mu}(\bold{S}^{-1}\bold{r}), \label{eq:lpb_q_c}
\end{align}
\label{eq:lpb_q}
\end{subequations}
where $\bold{S}$ and $\bar{\bold{q}}$ are defined in the same way as in Eqs.~\ref{eq:isdf_sym}. 

Eqs.~\ref{eq:local_pb} and~\ref{eq:lpb_q_c} provide a prescription for constructing local two-particle quantities in $\mathcal{C}$ directly from the intermediate objects expanded in the ISDF auxiliary basis. For instance, the local screened interactions in $\mathcal{C}$ read: 
\begin{subequations}
\begin{align}
&W_{\alpha\beta\gamma\delta}(i\Omega_{n}) = \frac{1}{N_{q}}\sum_{\bold{q}}W^{\bold{q}}_{\alpha\beta\gamma\delta}(i\Omega_{n}) \\
&= \frac{1}{N_{q}}\sum_{\bold{q}}\int d\bold{r}\int d\bold{r}' \Psi^{q*}_{\beta\alpha}(\bold{r})W(\bold{r},\bold{r}'; i\Omega_{n})\Psi^{\bold{q}}_{\gamma\delta}(\bold{r}') \\ 
&= \frac{1}{N_{q}}\sum_{\bold{q}}\sum_{\mu\nu}\Psi^{\bold{q}*}_{\beta\alpha}(\bold{S}\bold{r}_{\mu})W^{\bold{\bar{q}}}_{\mu\nu}(i\Omega_{n})\Psi^{\bold{q}}_{\gamma\delta}(\bold{S}\bold{r}_{\nu}), 
\end{align}
\end{subequations}
where $\bold{W}^{\bar{\bold{q}}}_{\mu\nu}(i\Omega_{n})$ is defined in Eq.~\ref{eq:ISDF_W_dyson}.  

\subsection{ISDF-cRPA \label{subsec:isdf_crpa}}
cRPA computes the effective Coulomb interactions for the subspace $\mathcal{C}$ within RPA with a modified irreducible polarizability. 
Given the substantial similarity with the $GW$ equations, the method can be easily incorporated into the ISDF formalism using an existing ISDF-$GW$ implementation.

In cRPA, the modified irreducible polarizability $\boldsymbol{\Pi}^{\bold{\bar{q}}}_{r}$ is defined as:
\begin{align}
\Pi_{r}^{\bold{\bar{q}}}(\bold{r},\bold{r}'; \tau) = \Pi^{\bold{\bar{q}}}(\bold{r},\bold{r}';\tau) - \Pi_{C}^{\bold{\bar{q}}}(\bold{r},\bold{r}';\tau).  
\label{Eq:chi_split_cRPA}
\end{align}
The particle-hole excitations within the correlated subspace $\boldsymbol{\Pi}^{\bold{\bar{q}}}_{C}$ are subtracted from the $GW$ polarizability $\boldsymbol{\Pi}^{\bold{\bar{q}}}$. 
$\boldsymbol{\Pi}^{\bold{\bar{q}}}_{C}$ is evaluated as:
\begin{align}
\Pi_{C}^{\bold{\bar{q}}}(\textbf{r}, \textbf{r}'; \tau) = \frac{1}{N_{k}}\sum_{\bold{k}}G_{C}^{\bold{k}}(\textbf{r}, \textbf{r}'; \tau)G_{C}^{\bold{k-\bar{q}}}(\textbf{r}', \textbf{r}; -\tau)
\end{align}
where $\bold{G}_{C}^{\bold{k}}$ is the Green's function projected from $\mathcal{C}$:
\begin{align}
G_{C}^{\bold{k}}(\bold{r},\bold{r}';\tau) = \sum_{\alpha\beta}w^{\bold{k}}_{\alpha}(\bold{r})G^{\bold{k}}_{\alpha\beta}(\tau)w^{\bold{k}*}_{\beta}(\bold{r}'). 
\end{align}

In the context of ISDF, we only need to evaluate $\boldsymbol{\Pi}^{\bar{\bold{q}}}_{r}$ at the interpolating points $\{\bold{r}_{\mu}\}$ and replace the $GW$ polarizability in Eq.~\ref{eq:ISDF_W_dyson} with $\boldsymbol{\Pi}^{\bar{\bold{q}}}_{r}$. 
The Dyson equation for the cRPA screened interactions $\bold{U}^{\bar{\bold{q}}}(i\Omega_{n})$ remains exactly the same:
\begin{align}
\bold{U}^{\bold{\bar{q}}}(i\Omega_{n}) = [\bold{I} - \bold{V}^{\bold{\bar{q}}}\boldsymbol{\Pi}^{\bar{\bold{q}}}_{r}(i\Omega_{n})]^{-1}\bold{V}^{\bold{\bar{q}}}. 
\label{eq:crpa_dyson}
\end{align}
The local cRPA screened interaction can be obtained via Eqs.~\ref{eq:lpb_q_c}:
\begin{subequations}
\begin{align}
&U_{\alpha\beta\gamma\delta}(\tau) = \frac{1}{N_{q}}\sum_{\bold{q}}U^{\bold{q}}_{\alpha\beta\gamma\delta}(\tau) \\
&= \frac{1}{N_{q}}\sum_{\bold{q}}\int d\bold{r}\int d\bold{r}' \Psi^{q*}_{\beta\alpha}(\bold{r})u(\bold{r},\bold{r}'; \tau)\Psi^{\bold{q}}_{\gamma\delta}(\bold{r}') \\ 
&= \frac{1}{N_{q}}\sum_{\bold{q}}\sum_{\mu\nu}\Psi^{\bold{q}*}_{\beta\alpha}(\bold{S}\bold{r}_{\mu})u^{\bold{\bar{q}}}_{\mu\nu}(\tau)\Psi^{\bold{q}}_{\gamma\delta}(\bold{S}\bold{r}_{\nu}), 
\end{align}
\label{eq:crpa_loc}
\end{subequations}
where $\bold{S}$ and $\bar{\bold{q}}$ are defined in the same way as in Eqs.~\ref{eq:isdf_sym}.

\subsection{Self-consistency and quasiparticle approximation\label{subsec:sc}}
As the simplest approximation to Hedin's equations, $GW$ requires full self-consistency between the Green's function and the self-energy accross the entire frequency spectrum. 
In the following, we will denote this method as ISDF-sc$GW$. 
Once the $GW$ self-energy is computed, a self-consistency iteration of ISDF-sc$GW$ is closed by the Dyson equation for the interacting Green's function in a single-particle basis:
\begin{align}
\bold{G}^{\bold{k}}(i\omega_{n}) = [(i\omega_{n}+\mu)\bold{I} - \bold{H}^{k}_{0} - \bold{J}^{\bold{k}} - \bold{\Sigma}^{\bold{k}}(i\omega_{n})]^{-1}, 
\label{eq:G_dyson}
\end{align}
where $\omega_{n} = (2n+1)\pi/\beta$ ($n\in \mathcal{Z}$) are fermionic Matsubara frequencies, $\mu$ is the chemical potential ensuring the correct total number of electrons, $\bold{H}^{\bold{k}}_{0}$ is the non-interacting Hamiltonian, $\bold{J}^{\bold{k}}$ is the Coulomb (Hartree) potential, and $\bold{\Sigma}^{\bold{k}}(i\omega_{n})$ is the $GW$ self-energy defined in Eqs.~\ref{eq:GW_sym}.  
Since ISDF-sc$GW$ does not rely on a quasiparticle approximation, all the off-diagonal elements of the self-energy are explicitly evaluated, and the inverse is computed through matrix inversion for a given $\bold{k}$ and $i\omega_{n}$. 
Subsequently, the updated interacting Green's function is iteratively fed back into the ISDF-sc$GW$ equations for $\Pi$, $W$, and $\Sigma$ until self-consistency is achieved. 

In cases where the initial guess (e.g., the DFT solution) is far from the converged solution, common techniques such as the direct inversion in the iterative subspace (DIIS)~\cite{Anderson_acceleration_1965,DIIS_Pulay_1980,DIIS_Pulay_1982,Anderson_acceleration_Walker2011,DIIS_Pavel_2022} can be employed to stabilize and expedite the convergence. 

Alternatively, the one-shot variant of $GW$, referred to as ISDF-$G_{0}W_{0}$, involves taking the $GW$ self-energy calculated from the first iteration based on a DFT non-interacting Hamiltonian and solving the following quasiparticle equation:
\begin{align}
\epsilon^{G_{0}W_{0}}_{n\bold{k}} = \epsilon^{\mathrm{KS}}_{n\bold{k}} - \mu
+ \mathrm{Re}\big[\Sigma^{\bold{k}}_{nn}(\epsilon^{G_{0}W_{0}}_{n\bold{k}})
- (V^{\mathrm{XC}})^{\bold{k}}_{nn} \big]. 
\label{eq:qp_eqn}
\end{align}
Here, $\epsilon^{\mathrm{KS}}_{n\bold{k}}$ is the $n$-th KS orbital energy at a crystal momentum $\bold{k}$, and $V^{\mathrm{XC}}$ is the exchange-correlation potential obtained from the starting DFT solution. 
In practice, Eq.~\ref{eq:qp_eqn} is solved either self-consistently or using a linear approximation around $\epsilon^{\mathrm{KS}}_{n\bold{k}}$. 
In this work, we analytically continued the $GW$ self-energy to the real-frequency axis using a Pad\'{e} approximation~\cite{Pade_Vidberg1977} and then solved Eq.~\ref{eq:qp_eqn} self-consistently via the bisection method.

From a computational standpoint, the additional cost of ISDF-sc$GW$ is directly proportional to the number of iterations required for self-consistency. 
In terms of computational cost per iteration, the two methods exhibit very similar complexities, given that both the Dyson equation and the Fourier transforms on the imaginary axis can be efficiently computed using the sparse sampling technique~\cite{sparse_sampling_Jia_2020}. 
In addition to the distinctions between the Dyson equation and the quasiparticle equation, ISDF-$G_{0}W_{0}$ can be slightly faster, as only the diagonals of the self-energy are required for updating the quasiparticle energies.

\section{Computational Details\label{sec:comp_detail}}
The starting points for our many-body calculations, including ISDF-$G_{0}W_{0}$, ISDF-sc$GW$, and ISDF-cRPA, are the DFT solutions calculated using \texttt{Quantum Espresso}~\cite{QE_Giannozzi2009,QE_Giannozzi2017,QE_Giannozzi2020}. 
In all cases, the DFT calculations employ the Perdew-Burke-Ernzerhof (PBE) exchange-correlation functional~\cite{PBE_Perdew1996} on a $14\times14\times14$ $\Gamma$-centered Monkhorst-Pack grid. 
Core electrons are described by either the correlated consistent effective core potentials (ccECPs)~\cite{ccecp_2ndrow_2017,ccecp_tm_2018,ccecp_4s4p_2019,ccecp_tm_2022}, optimized based on couple cluster theories, or the SG15 Optimized Norm-Conserving Vanderbilt (ONCV) pseudopotentials~\cite{ONCVPP_Hamann2013,SG15ONCV_Schlipf2015}, optimized for the PBE functional. 
Semi-core and valence electrons are treated explicitly and consistently throughout the DFT and many-body calculations. 
The energy cutoffs are determined to ensure that errors in DFT energies are less than $10^{-3}$ \emph{a.u.} per atom, and these values are consistently applied in both DFT and many-body calculations. 
Finally, the resulting KS orbitals, up to a truncated number $N_{\mathrm{orb}}$, are then taken as the single-particle Bloch basis for the subsequent many-body calculations. 

The many-body calculations are exclusively conducted on the imaginary axis at an inverse temperature of $\beta=1000$ \emph{a.u.} ($T \approx 316$ K). 
Dynamic quantities, including both fermionic and bosonic functions, are expanded into the intermediate representation (IR)~\cite{IR_Hiroshi_2017} with sparse sampling on both the imaginary-time and Matsubara frequency axes~\cite{sparse_sampling_Jia_2020}. 
Both the IR basis and the sampling points are generated using the \texttt{sparse-ir}~\cite{spare_ir_Markus2023} open-source software package.

Given the increased computational demand compared to DFT, the many-body calculations are performed on smaller $\bold{k}$-meshes. 
The typical dimensions range between $4\times4\times4$ and $9\times9\times9$, depending on the sizes of the unit cells. 
To facilitate convergence to the thermodynamic limit, we apply the finite-size correction scheme of Gygi and Baldereschi~\cite{Gygi_corrections1986} to quantities exhibiting integrable divergence, including the $GW$ self-energy and the local bare/screened Coulomb interactions in cRPA. 

\section{Results \label{sec:results}}
In this section, we present the results of our implementations of ISDF-$G_{0}W_{0}$, ISDF-sc$GW$, and ISDF-cRPA. 
We begin by validating the accuracy of the symmetry-adapted ISDF through an analysis of the convergence of the ISDF auxiliary basis in Sec.~\ref{subsec:thc_conv}. 
Following that, we present a benchmark against the published literature for selected semiconductors and insulators in Sec.~\ref{subsec:g0w0_band_gaps} and~\ref{subsec:scgw_band_gaps}. 
To showcase the capability of ISDF in large-scale many-body calculations, we combine ISDF-$G_{0}W_{0}$ and ISDF-cRPA to investigate the many-body states of carbon dimer defects in hexagonal boron nitride supercells in Sec.~\ref{subsec:hBN}. 
Finally, we provide a complexity analysis of our implementation in Sec.~\ref{subsec:complexity}. 

\subsection{Convergence of ISDF auxiliary basis \label{subsec:thc_conv}}
\begin{figure}[tbh!]
\begin{center}
\includegraphics[width=0.48\textwidth]{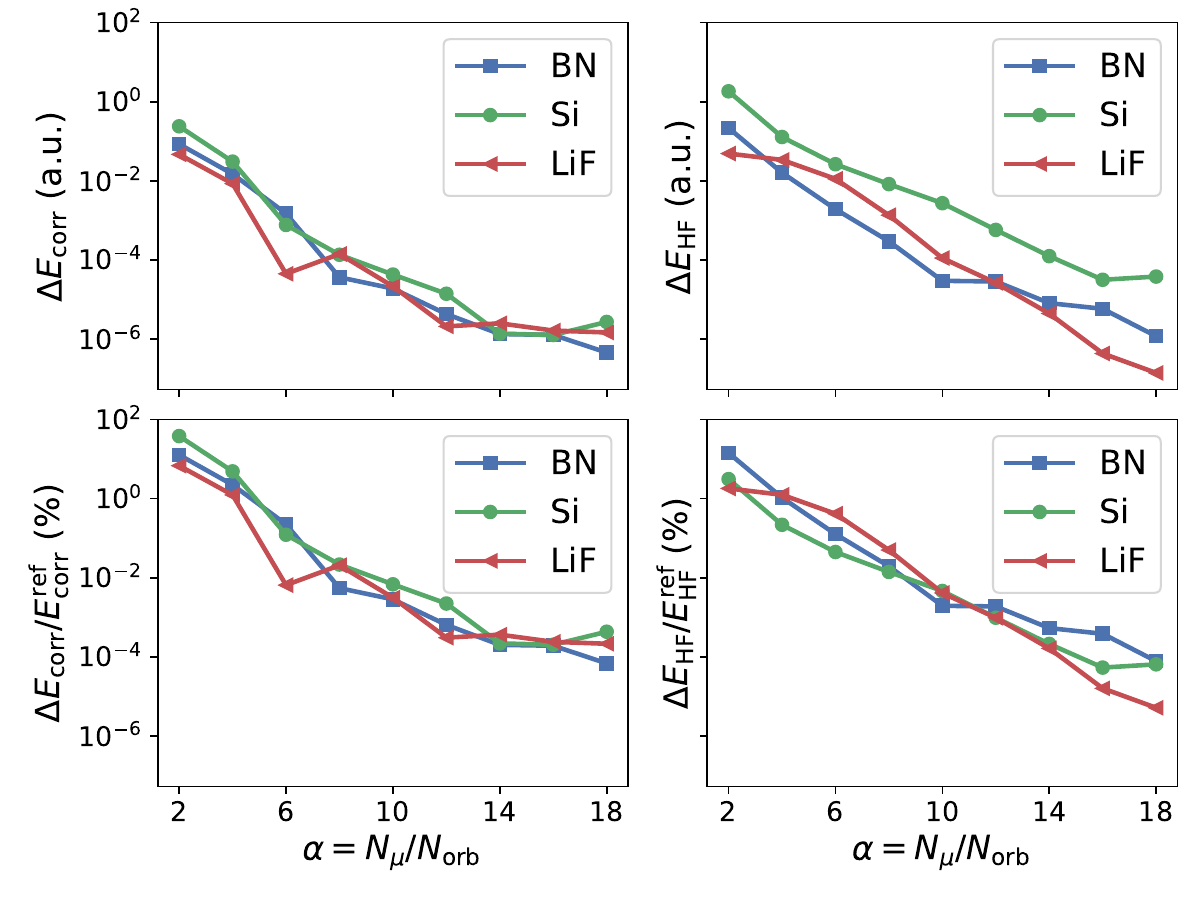} 
\caption{Absolute (top row) and relative errors (bottom row) of HF and correlation energies per atom for Si, BN and LiF with respect to the sizes of the auxiliary bases ($N_{\mu} = \alpha N_{\mathrm{orb}}$). }
\label{fig:thc_conv}
\end{center}
\end{figure}

As demonstrated in our previous work~\cite{THC-RPA_CNY2023}, both the ERIs and the RPA energy within the $\bold{q}$-resolved ISDF formalism exhibit exponential convergence with respect to the metric $\alpha = N_{\mu}/N_{\mathrm{orb}}$, where $N_{\mu}$ is characterized as a multiple of $N_{\mathrm{orb}}$. 
Moreover, it is observed that comparable accuracy can be attained at the same $\alpha$, regardless of the number of virtual orbitals, the number of $\bold{k}$-points, and the size of a unit cell.

In this section, we analyze the convergence of the $GW$ self-energy concerning $\alpha$ within the framework of symmetry-adapted ISDF. Fig.~\ref{fig:thc_conv} illustrates the errors of the HF and correlation energies for Si, BN, and LiF from the first iteration of ISDF-sc$GW$ as a function of $\alpha$. 
The HF and correlation energies are computed using the Galitskii-Migdal formula which directly reflects the accuracy of the static and the dynamic $GW$ self-energy, respectively.

For reference data, we utilize the $GW$ results calculated through our in-house implementation based on Cholesky decomposed ERIs without considering space-group symmetries. 
All calculations employ ccECPs on a $2\times2\times2$ $\Gamma$-centered Monkhorst-Pack $\bold{k}$-mesh, with the number of KS states ($N_{\mathrm{orb}}$) chosen to be 8 times the total number of electrons per unit cell. 
The use of a small $\bold{k}$-mesh is intended to alleviate the computational cost of the Cholesky-based $GW$, which exhibits quadratic scaling with respect to $N_{k}$. 
Since the convergence of the RPA energy concerning $\alpha$ is found to be independent of the choice of $\bold{k}$-meshes~\cite{THC-RPA_CNY2023}, we anticipate similar behavior for the $GW$ self-energy.

Despite the incorporation of space-group symmetries, our symmetry-adapted ISDF exhibits fairly consistent convergence, aligning with the findings in Ref.~\onlinecite{THC-RPA_CNY2023}. As $\alpha$ increases, the three selected systems demonstrate similar exponential convergence, even though their single-particle bases differ in size. This observation reflects the fact that the rank of the pair densities grows linearly with respect to $N_{\mathrm{orb}}$ for any generic single-particle basis.
In terms of comparing HF and correlation energies, their convergences are notably similar, directly stemming from the equal treatment of occupied and virtual orbitals in the ISDF procedures.

Quantitatively, the absolute errors in the HF energy for Si are systematically larger, primarily due to the presence of semi-core electrons ($2s$ and $2p$ for Si) in the ccECP, leading to a larger magnitude in the HF energy. 
More consistent convergences across the three systems become apparent when examining the relative errors in the bottom row of Fig.~\ref{fig:thc_conv}. 
Conversely, the convergence of the $GW$ correlation energy remains consistently stable across the three systems, irrespective of the presence of semi-core electrons. 
It is noteworthy that the necessity for a larger $N_{\mu}$ for the HF energy aligns with a previous finding in Ref.~\onlinecite{THC_MP3_Joonho2020}, where ISDF is applied to all-electron molecular calculations. 
Given that the HF potential solely depends on occupied orbitals, one potential approach to address this issue could involve constructing a customized ISDF decomposition for HF in which all virtual orbitals are excluded. 

As proposed in Ref.~\onlinecite{THC-RPA_CNY2023}, it is suggested that $\alpha=8$ is adequate to achieve chemical accuracy when using the ONCV pseudopotentials. However, for the instances involving significantly harder ccECPs, our findings indicate that a slightly larger ISDF auxiliary basis ($\alpha=10\sim12$) is necessary to attain chemical accuracy for both HF and correlated methods. 

\subsection{\texorpdfstring{$G_{0}W_{0}$}{G0W0} band gaps\label{subsec:g0w0_band_gaps}}
\begin{table*}[bth]
\begin{center}
\begin{tabular}{c|c|c|c|c|c}
\hline
\hline
System & lattice parameter (\AA) & $E_{\mathrm{cutoff}}$ (Ha) & $\bold{k}$-mesh & $N_{\mathrm{KS}}$($G_{0}W_{0}$/sc$GW$) & $\alpha=N_{\mathrm{aux}}/N_{\mathrm{KS}}$ \\
\hline
Si & 5.43~\cite{PhysRevB.69.075102} & 450 & 8x8x8 & 240 & 10 \\
SiC & 4.35~\cite{Yu2010} & 450 & 8x8x8 &  240 & 10 \\
AlP & 5.451~\cite{Sze2006} & 350 & 8x8x8 & 240 & 10\\
ZnS & 5.420~\cite{Sze2006} & 225 & 5x5x5 & 1300/900  & 10 \\
C & 3.567~\cite{Yu2010} & 140 & 9x9x9 & 240 & 10 \\
BN & 3.615~\cite{Sze2006} & 190 & 8x8x8 & 160 & 10 \\
MgO & 4.213~\cite{MgO_lattice} & 275 & 8x8x8 & 240 & 10 \\ 
LiF & 4.030 & 275 & 8x8x8 & 500 & 10 \\ 
\hline
\end{tabular}
\caption{Details of the computational parameters for ISDF-$G_{0}W_{0}$ and ISDF-sc$GW$. 
\label{tab:comp_details_ccecp}}
\end{center}
\end{table*}

\begin{table*}[bth]
\begin{center}
\begin{tabular}{c|c|cccc|c}
\hline
\hline
System & PBE & & $G_{0}W_{0}$@PBE & & & Expt\\
& & this work & LAPW~\cite{LAPW_G0W0_basis_conv} & GTO~\cite{G0W0_Zhu2021} & PAW~\cite{VASP_evGW_2007} & \\
\hline
Si & 0.61 &  1.04 &1.12 &1.08 & 1.12 & 1.17~\cite{Kittel2004} \\
SiC & 1.36 & 2.35 & 2.38 & 2.42 & 2.27 & 2.40~\cite{Yu2010} \\
AlP & 1.52 & 2.41  & 2.37 & 2.41& 2.44 & 2.45~\cite{Sze1981} \\
ZnS & 2.06 & 3.40 & 3.35 & 3.63& 3.29 & 3.91~\cite{Kittel2004} \\ 
C & 4.16 & 5.63 & 5.69 & 5.52 & 5.50 & 5.48~\cite{Yu2010} \\ 
BN & 4.57 & 6.52 & 6.36 & 6.41 & 6.10 & 6.40~\cite{madelung2004semiconductors} \\
MgO & 4.81 &  7.45 & 7.52 & 7.43 & 7.25 & 7.83~\cite{MgO_Expt_1973} \\
LiF & 9.10 & 13.76 & 14.27 & & 13.27 & 14.2~\cite{LiF_expt_1976} \\
\hline
\hline
\end{tabular}
\caption{Band gaps (eV) for selected semiconductors and insulators calculated using ISDF-$G_{0}W_{0}$, compared with theoretical and experimental literature. \label{tab:g0w0_band_gaps_ccecp}}
\end{center}
\end{table*}
We present the $G_{0}W_{0}$ band gaps for the selected semiconductors and insulators using the symmetry-adapted ISDF formalism. 
To facilitate quantitative comparison with published all-electron calculations, we employ hard ccECPs~\cite{ccecp_2ndrow_2017,ccecp_tm_2018,ccecp_4s4p_2019,ccecp_tm_2022} wherein semi-core electrons are treated explicitly. Table.~\ref{tab:comp_details_ccecp} provides a summary of computational details for different systems, including energy cutoffs ($E_{\mathrm{cutoff}}$), the number of KS orbitals ($N_{\mathrm{orb}}$), and the sizes of $\bold{k}$-meshes.

The convergence with respect to $N_{\mathrm{orb}}$ varies among different systems depending on the chemical properties of atomic species. 
Transition metal elements, for instance, typically exhibit a need for much larger $N_{\mathrm{orb}}$ in the KS basis due to strong electron correlations. 
To estimate basis set truncation errors, we systematically increase $N_{\mathrm{orb}}$ and then extrapolate the band gaps to the infinite basis set limit $E_{g}(N_{\mathrm{orb}}\rightarrow\infty)$ based on the formula $E_{g}(N_{\mathrm{orb}}) = a/N_{\mathrm{orb}} + b$.
Given the computational cost in the large $N_{\mathrm{orb}}$ limit, the extrapolation is performed on coarse $\bold{k}$-meshes ranging in size from $3\times3\times3$ to $5\times5\times5$. The resulting basis set corrections are then directly applied to the finer $\bold{k}$-meshes listed in Table~\ref{tab:comp_details_ccecp}, assuming the corrections are $\bold{k}$-independent. 
For ISDF-$G_{0}W_{0}$ band gaps, we find that the corresponding basis set corrections are consistently below 0.15 eV across all selected systems. The complete data on basis set convergence is provided in Supporting Information. 

Table~\ref{tab:g0w0_band_gaps_ccecp} shows the band gaps of selected systems calculated using PBE and ISDF-$G_{0}W_{0}$, referred to as "this work." 
To facilitate comparison, predictions from other $G_{0}W_{0}$ literature~\cite{LAPW_G0W0_basis_conv,G0W0_Zhu2021,VASP_evGW_2007}, along with experimental data~\cite{Kittel2004,Yu2010,Sze1981,madelung2004semiconductors,MgO_Expt_1973,LiF_expt_1976}, are also presented.

Starting from the PBE effective Hamiltonians, the many-body corrections from the $G_{0}W_{0}$ self-energy induce band gap widening, aligning the predicted band gaps more closely with the experimental references. Except diamond, ISDF-$G_{0}W_{0}$ only slightly underestimates the experimental band gaps by $2-13\%$. 
These favorable agreements with the experimental data are often attributed to the fortuitous error cancellation between the lack of self-consistency and vertex corrections, as reported in various literature~\cite{GW_vertex_Gruneis_2014,scGW_w_vertex_Andrey_2016,scGW_also_vertex_Andrey_2017}. 
However, since the dependence on the starting mean-field solution could be on the same magnitude as the vertex corrections~\cite{scGW_vs_vertex_wen2023}, achieving systematic improvement on top of the $G_{0}W_{0}$ solutions becomes a non-trivial problem.

We also compare our results of ISDF-$G_{0}W_{0}$ with all-electron calculations using different single-particle bases, including linearized augmented plane waves (LAPWs), projector augmented waves (PAWs), and Gaussian-type orbitals (GTOs). 
This comparison not only helps establish the accuracy of symmetry-adapted ISDF but also serves as a stringent test for the frozen-core approximation in the presence of the core-core and core-valence correlations originating from many-body theories.

For all the systems considered in this work, we found that the differences between our ISDF-$G_{0}W_{0}$ band gaps and the existing all-electron calculations~\cite{LAPW_G0W0_basis_conv,G0W0_Zhu2021,scGW_VASP_2018} are well below the deviation compared to the experimental values. The quantitative agreement suggests that core-core and core-valence correlations on the valence properties are mostly captured in the construction of ccECPs.

In the ccECPs employed in the present work~\cite{ccecp_2ndrow_2017,ccecp_tm_2018,ccecp_4s4p_2019,ccecp_tm_2022}, these types of correlations are accounted for by optimizing the effective potentials using coupled cluster theories at the atomic level. 
This ensures much better transferability to many-body calculations at the price of a much higher energy cutoff compared to a typical norm-conserving pseudopotential optimized for DFT. 
Our data suggests that quantitative agreements between pseudopotential and all-electron calculations are achievable with the help of hard ccECPs. 

In practice, the high energy cutoffs in these hard ccECPs result in FFT grids of substantial size, rendering the space-time formalism computationally infeasible. 
In contrast, the ISDF formalism is dramatically more efficient since the ccECP cases fall into the limit where $N_{\mu}$ is considerably smaller than $N_{r}$. 
The problematic FFT grids arise only in the ISDF decomposition of ERIs, and the evaluations of all dynamic quantities are conducted on the dimension of $N_{\mu}$ rather than $N_{r}$.

\subsection{\texorpdfstring{sc$GW$}{scGW} band gaps\label{subsec:scgw_band_gaps}}
\begin{table*}[bth]
\begin{center}
\begin{tabular}{c|c|cccc|c}
\hline
\hline
System & PBE & & sc$GW$ & & & Expt\\
& & this work & LAPW~\cite{scGW_also_vertex_Andrey_2017} & GTO~\cite{scGW_CNY2022} & PAW~\cite{scGW_VASP_2018} \\
\hline
Si & 0.61 & 1.49 & 1.55 & 1.50  & 2.18 & 1.17~\cite{Kittel2004} \\
SiC & 1.36 & 2.98 & 2.89 & 2.95 & 3.29 & 2.40~\cite{Yu2010} \\
AlP & 1.52 & 3.03 & 2.84 & 2.90 & 3.20 & 2.45~\cite{Sze1981} \\
ZnS &  2.06 & 4.51 & 4.28 & 4.50 & 4.68 & 3.91~\cite{Kittel2004} \\ 
C & 4.16 & 6.18 & & & 6.41 & 5.48~\cite{Yu2010} \\ 
BN & 4.57 & 7.38 & 7.06 & 7.17 & 7.67 & 6.40~\cite{madelung2004semiconductors} \\
MgO & 4.81 & 9.06 & 9.31 & 9.29 & 9.53 & 7.83~\cite{MgO_Expt_1973} \\
LiF & 9.10 & 15.87 & 16.30 & & & 14.2~\cite{LiF_expt_1976} \\
\hline
\hline
\end{tabular}
\caption{
Band gaps (eV) for selected semiconductors and insulators calculated using ISDF-sc$GW$, compared with theoretical and experimental literature. 
\label{tab:scgw_band_gaps_ccecp}}
\end{center}
\end{table*}
We now examine the results of fully self-consistent $GW$ employing the symmetry-adapted ISDF formalism. 
Given the full frequency dependence in the self-energy, the notion of quasiparticle energies is no longer applicable in sc$GW$. 
Instead, we compute $\bold{k}$-resolved spectral functions by analytically continuing the single-particle Green's functions. 
The band gaps are then defined as the peak-to-peak distance between the quasiparticle peaks. 

Table~\ref{tab:scgw_band_gaps_ccecp} presents the band gaps of selected semiconductors and insulators, calculated using PBE and ISDF-sc$GW$, alongside reference data from theoretical~\cite{scGW_also_vertex_Andrey_2017,scGW_VASP_2018,scGW_CNY2022} and experimental~\cite{Kittel2004,Yu2010,Sze1981,madelung2004semiconductors,MgO_Expt_1973,LiF_expt_1976} literature. Computational details for ISDF-sc$GW$, denoted as "this work", can be found in Table~\ref{tab:comp_details_ccecp}.

In comparison to $G_{0}W_{0}$, self-consistency induces further gap widening, leading to a systematic overestimation of the band gaps. The overestimation ranges from 12\% to 28\% for the studied systems, similar to findings in quasiparticle self-consistent $GW$ (QPGW)~\cite{LQSGW_Kutepov2017}. 
While the accuracy is slightly inferior to that of the one-shot $G_{0}W_{0}$, due the absence of starting-point dependence, the overestimation can be unambiguously attributed to the lack of vertex corrections in the $GW$ polarizability and the self-energy~\cite{scGW_w_vertex_Andrey_2016,scGW_also_vertex_Andrey_2017}. 
Another advantage of sc$GW$ is that the self-consistency loop can be executed exclusively on the imaginary axis with no need for analytical continuation. 
This is particularly useful in calculations involving orbitals far away from the Fermi level, which could potentially compromise the stability of self-consistency during the analytical continuation. 

When compared to the published theoretical literature, our data demonstrate quantitative agreements with the sc$GW$ results obtained from all-electron implementations based on LAPWs~\cite{scGW_also_vertex_Andrey_2017} and GTOs~\cite{scGW_CNY2022}. 
The differences are smaller than the deviations observed when compared to experimental values. 
This justifies the validity of the frozen-core approximation even in a method that does not rely on the quasiparticle approximation. 
Nevertheless, we would like to emphasize that the correlated consistent treatment in the construction of the ccECPs is essential for good transferability beyond mean-field calculations~\cite{ccecp_2ndrow_2017,ccecp_tm_2018,ccecp_4s4p_2019,ccecp_tm_2022}. 
In contrast, the sc$GW$ band gaps calculated using PAWs~\cite{scGW_VASP_2018} exhibit systematic overestimation compared to other implementations. 
Since errors arising from basis set truncation and finite-size effects are properly addressed in all these literature, the difference is potentially due to the inconsistent definitions of the band gap, as discussed in Ref.~\onlinecite{scGW_CNY2022}. 

\subsection{Carbon dimer in hexagonal boron nitride \label{subsec:hBN}}
To demonstrate the capability of the symmetry-adapted ISDF formalism in addressing large-scale systems, we integrate ISDF-$G_{0}W_{0}$ and ISDF-cRPA to investigate the many-body (MB) states of the neutral carbon dimer defect $\mathrm{C_{B}C_{N}}$ in hexagonal boron nitride (hBN). 
This process involves constructing a downfolded impurity Hamiltonian and subsequently solving it using a high-level impurity solver. 
It's important to note that the downfolding procedures for defect systems, based on an \emph{ab initio} many-body calculation, can be as computationally demanding as the impurity solver since it typically requires a large supercell to minimize interactions between defects and their periodic images. The presence of the defect also reduces the number of space group symmetries, posing an additional computational overhead. 

As the correlation effects from core electrons are expected to be small in the case of $\mathrm{C_{B}C_{N}}$, we utilize the softer SG15 ONCV pseudopotentials~\cite{ONCVPP_Hamann2013,SG15ONCV_Schlipf2015}. 
Atomic positions are relaxed for different supercells using PBE functional on a $2\times2\times2$ $\Gamma$-centered $\bold{k}$-mesh. 
For both DFT and many-body calculations, the kinetic energy cutoffs are set to 35 \emph{a.u.}, ensuring DFT energy convergence within $10^{-3}$ \emph{a.u.} per atom. 
The size of the ISDF auxiliary basis is chosen as $N_{\mu}=10N_{\mathrm{orb}}$, where HF and correlation energies are converged within the same threshold. 
Following the construction of the downfolded Hamiltonian, MB excitation energies are computed using exact diagonalization (ED) within the \texttt{PySCF} open-source package~\cite{PySCF_2020}.

We employ the downfolding procedures outlined in Refs.~\onlinecite{Bockstedte2018,PRB_Muechler2022,QDET_Sheng2022}. 
Starting from a PBE solution, we construct a localized basis that characterizes the $\mathrm{C_{B}C_{N}}$ defect states through Wannierization~\cite{wannier90_Mostofi2014}. 
The complete electronic structure is subsequently downfolded to the subspace spanned by the Wannier functions, resulting an effective impurity Hamiltonian: 
\begin{align}
\hat{H}^{G_{0}W_{0}+\mathrm{cRPA}}_{\mathrm{eff}} = \sum_{\alpha\beta}(&t^{G_{0}W_{0}}_{\alpha\beta}-\mu\delta_{\alpha\beta})\hat{c}^{\dag}_{\alpha}\hat{c}_{\beta} \\
&+ \frac{1}{2}\sum_{\alpha\beta\gamma\delta}v^{\mathrm{cRPA}}_{\alpha\beta\gamma\delta}\hat{c}^{\dag}_{\alpha}\hat{c}^{\dag}_{\gamma}\hat{c}_{\delta}\hat{c}_{\beta}, \nonumber
\label{eq:eff_hamilt}
\end{align}
in which the Greek letters denote the Wannier functions for the defects states, $\bold{t}^{G_{0}W_{0}}$ is the downfolded one-body Hamiltonian obtained from the ISDF-$G_{0}W_{0}$ band structure with the double counting (DC) contributions subtracted and $\bold{v}^{\mathrm{cRPA}}$ is the ISDF-cRPA screened interactions. 
In principle, there exists hybridization between the Wannier functions and the environment which can be described by adding interactions with a bath of non-interacting electrons. 
For this particular system, we find the hybridization function always remains negligible and, therefore, discard it. 
For comparison, we also compute the MB states of $\hat{H}^{\mathrm{PBE+cRPA}}_{\mathrm{eff}}$ using the procedures outlined in Ref.~\onlinecite{PRB_Muechler2022}. 
The explicit definitions of $\hat{H}^{G_{0}W_{0}\mathrm{+cRPA}}_{\mathrm{eff}}$ and $\hat{H}^{\mathrm{PBE+cRPA}}_{\mathrm{eff}}$ can be found in Supporting Information. 

\begin{figure}[tbh!]
\begin{center}
\includegraphics[width=0.30\textwidth]{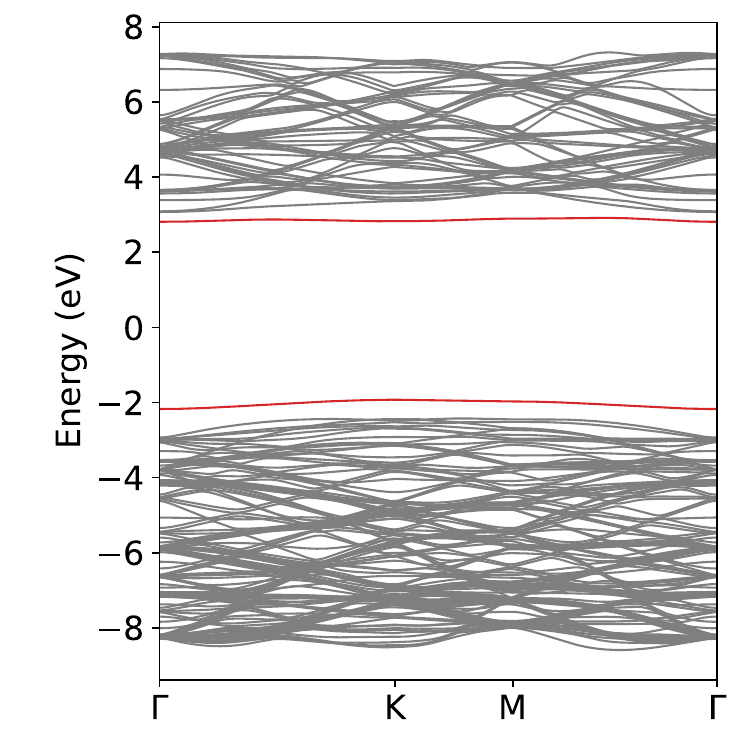} 
\caption{ISDF-$G_{0}W_{0}$@PBE band structure for $\mathrm{C_{B}C_{N}}$ in $4\times4\times2$ hBN supercells and a $4\times 4\times2$ Monkhorst-Pack $\bold{k}$-mesh. The red curves represent the bonding and anti-bonding states from the $p_{z}$ orbitals in $\mathrm{C_{B}C_{N}}$. }
\label{fig:hBN_bands}
\end{center}
\end{figure}

Fig.~\ref{fig:hBN_bands} displays the ISDF-$G_{0}W_{0}$@PBE band structure for $\mathrm{C_{B}C_{N}}$ in $4\times4\times2$ hBN supercells with a $4\times 4\times2$ Monkhorst-Pack $\bold{k}$-mesh. 
The introduction of a neutral carbon dimer by replacing a pair of boron and nitrogen atoms on the nearest-neighbor sites results in two defect states, denoted by the red curves, situated within the band gap of hBN. 
These defect states represent the bonding and anti-bonding states with carbon $p_{z}$ character, and their flatness reflects the zero-dimensional nature of $\mathrm{C_{B}C_{N}}$. 
In the Wannierization process, these two flat bands are specifically chosen to construct the Wannier functions. 

The ground state of $\mathrm{C_{B}C_{N}}$ is the singlet state, $^{1}A_{1}$,  where the bonding orbitals are fully occupied. 
At the PBE level, the band gaps for the defect and hBN at the $\Gamma$-point are 3.59 and 4.72 eV, respectively. 
In the subsequent $G_{0}W_{0}$ calculation, the many-body corrections increase the charged excitation gaps to 4.99 and 6.00 eV. 

\begin{figure}[tbh!]
\begin{center}
\includegraphics[width=0.48\textwidth]{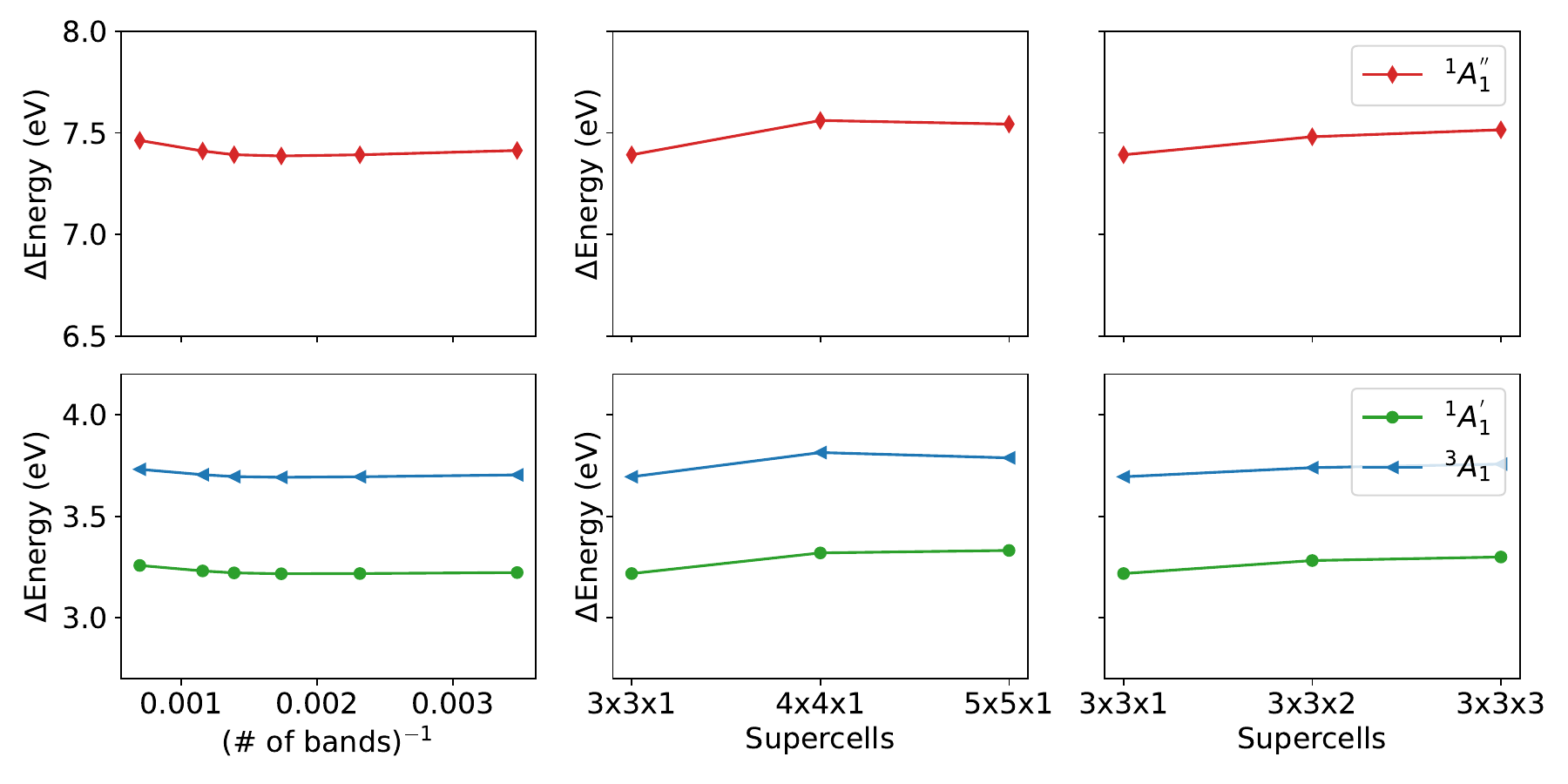} 
\caption{MB excitation energies (eV) of $\mathrm{C_{B}C_{N}}$ defect states with respect to the numbers of bands and the sizes of supercells along different directions.}
\label{fig:hBN_mb_conv}
\end{center}
\end{figure}
We now examine the convergence behavior of the MB excitation energies of $\mathrm{C_{B}C_{N}}$ concerning different parameters. 
In this context, the excitation energies are defined relative to the ground-state singlet $^{1}A_{1}$. 
The first excited state, $^{3}A_{1}$, represents the triplet state obtained by exciting one electron to the anti-bonding orbital. 
The subsequent excited state, $^{1}A^{'}_{1}$, is the open-shell singlet where both the bonding and the anti-bonding orbitals are singly occupied with opposite spins. 
Finally, the third excited state, $^{1}A^{''}_{1}$, corresponds to a closed-shell singlet with a fully occupied anti-bonding orbital. 

Fig.~\ref{fig:hBN_mb_conv} illustrates the $\mathrm{C_{B}C_{N}}$ excitation energies concerning the number of bands and the supercell dimensions. 
In contrast to the benchmark systems discussed in Sec.~\ref{subsec:g0w0_band_gaps} and~\ref{subsec:scgw_band_gaps}, the basis set convergence of $\mathrm{C_{B}C_{N}}$ is remarkably rapid. 
A value of $N_{\mathrm{orb}} = 3N_{\mathrm{e}}$ is already sufficient to attain the same accuracy in our ISDF setup. 

Concerning convergence with respect to supercell dimensions, we enlarge the supercell along the $xy$ and $z$ directions separately, perserving the hexagonal structure of the host material. 
Along the $xy$ direction, reasonably converged results are achieved with a quadruple unit cell.
By enlarging the supercell along the $z$ direction, the changes in $\mathrm{C_{B}C_{N}}$ excitation energies from a $3\times3\times2$ supercell to a $3\times3\times3$ supercell are less than $10^{-3}$ \emph{a.u.}. 
In summary, we conclude that a $4\times4\times2$ supercell is sufficiently large to spatially isolate the carbon dimer, thereby preserving its zero-dimensional nature. 

\begin{table}[bth]
\begin{center}
\begin{tabular}{c|ccc}
\hline
\hline
& ED@PBE+cRPA& ED@$G_{0}W_{0}$+cRPA & Expt \\
 & (Hartree DC) & & ~\cite{CBCN_GW_BSE_PRM2021} \\
\hline
$^{3}A_{1}$ & 3.52 & 3.34 & - \\
$^{1}A^{'}_{1}$  & 3.99 & 3.81 & 4.6\\ 
$^{1}A^{''}_{1}$  & 7.95 & 7.59 & - \\
\hline
\hline
\end{tabular}
\caption{MB excitation energies (eV) of $\mathrm{C_{B}C_{N}}$ calculated using two different downfolded Hamiltonians, compared with available experimental data~\cite{CBCN_GW_BSE_PRM2021}. 
\label{tab:dimer_mb_states}}
\end{center}
\end{table}

After establishing the zero-dimensional nature of $\mathrm{C_{B}C_{N}}$, we proceed to compare the $\mathrm{C_{B}C_{N}}$ excitation energies obtained from downfolded Hamiltonians based on PBE (referred to as ED@PBE+cRPA) and $G_{0}W_{0}$@PBE (referred to as ED@$G_{0}W_{0}$+cRPA) band structures. 
Since the two-electron Hamiltonians are computed using ISDF-cRPA in both cases, the distinction lies in the one-body Hamiltonian, which is predominantly influenced by single-particle energies and the DC corrections. 
Following the approach outlined in Refs.~\onlinecite{Bockstedte2018,PRB_Muechler2022}, the DC contribution in $\hat{H}^{\mathrm{PBE+cRPA}}_{\mathrm{eff}}$ is evaluated as the Hartree term, excluding the contribution from the exchange-correlation functional. 


Table~\ref{tab:dimer_mb_states} presents the $\mathrm{C_{B}C_{N}}$ excitation energies calculated using the ED for $\hat{H}^{\mathrm{PBE+cRPA}}_{\mathrm{eff}}$ and $\hat{H}^{G_{0}W_{0}\mathrm{+cRPA}}_{\mathrm{eff}}$. The distinctions in their MB states can be qualitatively understood through an analogy with a Hubbard dimer model in open boundary conditions. In the notation of our Wannier functions, the two sites correspond to the two C atoms, each bonded to different neighboring atoms (N and B atoms), resulting in an asymmetry in onsite potentials.
Despite the $G_{0}W_{0}$@PBE band gap being approximately 1.5 eV larger, the difference in their DC terms mostly compensates, resulting in only a slight deviation of -0.03 eV for the off-diagonals in $t_{\alpha\beta}$ from PBE to $G_{0}W_{0}$@PBE. 
Additionally, the asymmetry, defined as the difference between the onsite potentials of the two C atoms, decreases by 0.22 eV from PBE to $G_{0}W_{0}$@PBE. 
Together, the decrease in inter-site hopping and onsite asymmetry leads to smaller excitation energies in $\hat{H}^{G_{0}W_{0}\mathrm{+cRPA}}_{\mathrm{eff}}$.

Compared to the experiment value, the first singlet excitation energy coming from ED@$G_{0}W_{0}$+cRPA underestimates the experimental value by 0.8 eV, inferior to the prediction from ED@PBE+cRPA. 
However, it is essential to note that there is a missing DC contribution arising from the exchange-correlation potential. 
Thus, we refrain from conclusively asserting that ED@PBE+cRPA is systematically more accurate than ED@$G_{0}W_{0}$+cRPA. 
Conversely, the DC contribution in $GW$ embedding is rigorously defined in terms of the self-energy diagrams. 
A more comprehentive benchmark for comparing the two downfolded Hamiltonians is necessary and would be an interesting topic in the future. 

The underestimation of the first singlet excitation energy in $\hat{H}^{G_{0}W_{0}\mathrm{+cRPA}}_{\mathrm{eff}}$ can be attributed to the limitations of cRPA and its static approximation. 
Analyses of the Hubbard model~\cite{cRPA_Hiroshi2015,limitation_cRPA_Honerkamp2018} suggest that cRPA tends to overestimate screening due to the absence of cancellation effects between cRPA and non-cRPA diagrams. 
This overestimation in screening effects results in effective interactions that are too small for the downfolded Hamiltonian. 
Simultaneously, as demonstrated in Ref.~\onlinecite{Uw_PRB_Casula2012,Uw_PRL_Casula2012}, the frequency-dependence of screened interactions introduces additional renormalizations of the one-body Hamiltonian. 
Effectively, a model that relies on instantaneous interactions will require larger interactions to account for these additional renormalizations.

Another source of errors arises from the tendency of $G_{0}W_{0}$@PBE to underestimate quasiparticle band gaps, thereby influencing the downfolded one-body Hamiltonian. 
Downfolding based on $G_{0}W_{0}$ calculations starting from a hybrid functional can effectively increase excitation energies. 
Lastly, it is noteworthy that the hybridization between $\mathrm{C_{B}C_{N}}$ and the host material is found to be negligible across all different supercell sizes. 
Consequently, we conclude that the addition of hybridization functions in the downfolded Hamiltonian will not significantly impact the quantitative results.

\subsection{Complexity analysis\label{subsec:complexity}}
\begin{figure}[tbh!]
\begin{center}
\includegraphics[width=0.40\textwidth]{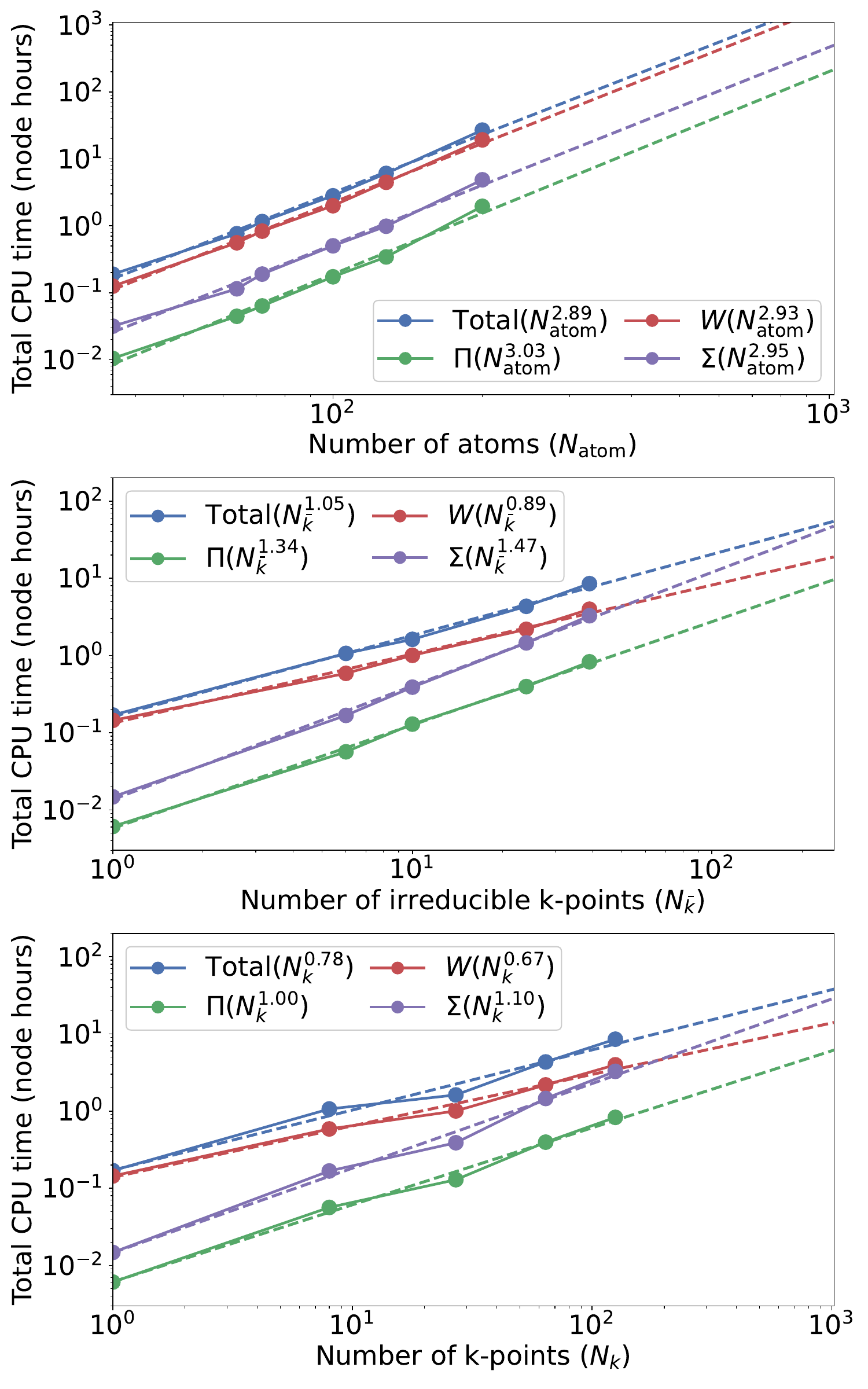} 
\caption{Total CPU time (node hours) for one iteration of ISDF-sc$GW$ with $\alpha=10$. Top panel: $\mathrm{C_{B}C_{N}}$ in hBN supercells with a $2\times2\times2$ $\bold{k}$-mesh. Middle panel: $\mathrm{C_{B}C_{N}}$ in $3\times3\times2$ hBN supercells with different number of irreducible $\bold{k}$-points. Bottom panel: $\mathrm{C_{B}C_{N}}$ in $3\times3\times2$ hBN supercells with different number of $\bold{k}$-points. }
\label{fig:cputime}
\end{center}
\end{figure}
Finally, we turn to investigate the computational complexity of the ISDF-$GW$ equations concerning the number of atoms in a unit cell ($N_{\mathrm{atom}}$) and the $\bold{k}$-point sampling. 
Since the computational bottlenecks of ISDF-$G_{0}W_{0}$, ISDF-sc$GW$, and ISDF-cRPA all arise from these equations, the analysis in this section can be directly applied to these methods. 
The analysis is conducted for $\mathrm{C_{B}C_{N}}$ defects embedded in hBN supercells, using the same numerical setups as in Sec.~\ref{subsec:hBN}. 
$N_{\mathrm{orb}}$ is set to three times the total number of electrons, with $N_{\mu} = 3N_{\mathrm{orb}}$ for all data points.

The top panel of Fig.~\ref{fig:cputime} illustrates the total CPU node hours concerning $N_{\mathrm{atom}}$ for one iteration of ISDF-sc$GW$, including the evaluation of $\Pi$, the Dyson equation for $W$, and the evaluation of $\Sigma$. 
For consistency, we use a $2\times2\times2$ Monkhorst-Pack $\bold{k}$-mesh for all the supercell calculations here. 
As we systematically enlarge the sizes of hBN supercells from $3\times3\times1$ ($N_{\mathrm{atom}}=36$) to $5\times5\times2$ ($N_{\mathrm{atom}}=200$), consistent cubic scalings are observed in all the steps of ISDF-$GW$. 
As expected, the Dyson equation for $W$ is the most computationally expensive due to the large prefactor coming from the cubic scaling with respect to $N_{\mu}$. 
The evaluation of $\Pi$ is less computationally expensive than the evaluation of $\Sigma$ due to the particle-hole symmetry presented in the bosonic quantities. 

Regarding the computational time with respect to the $\bold{k}$-point sampling, we investigate $\mathrm{C_{B}C_{N}}$ defects embedded in $3\times3\times2$ hBN supercells with $n\times n\times n$ Monkhorst-Pack $\bold{k}$-meshes ($n=1\sim5$). 
In the presence of space-group symmetries, the numbers of irreducible $\bold{k}$-points $(N_{\bar{k}})$ are 1, 6, 10, 24, and 39. 
As depicted in the middle panel of Fig.~\ref{fig:cputime}, the three components of ISDF-$GW$ display somewhat different complexities. 
Except for the Dyson equation for $W$, both the evaluations of $\Pi$ and $\Sigma$ deviate from a linear scaling with respect to $N_{\bar{k}}$. 
As $N_{\bar{k}}$ increases, the computing time of these two steps will eventually surpass the evaluation of $W$, resulting in the overall complexity deviating further from $O(N_{\bar{k}})$.

We would like to emphasize that these deviations from $O(N_{\bar{k}})$ are entirely expected since both $\Pi$ and $\Sigma$ require the evaluation of $G^{\bold{k}}(\bold{r}_{\mu}, \bold{r}_{\nu}; \tau)$ outside the IBZ, regardless of whether space-group symmetries are activated or not.
A more reasonable analysis would be comparing the computing time at different total numbers of $\bold{k}$-points ($N_{k}$), as shown in the bottom panel of Fig.~\ref{fig:cputime} where both $\Pi$ and $\Sigma$ can be evaluated at the cost of $O(N_{k})$.

\section{Conclusion \label{sec:conclusion}}
In conclusion, we have introduced efficient algorithms for solving the $GW$ and cRPA equations within a generic Bloch basis. 
These algorithms exhibit favorable scaling behaviors, demonstrating a cubic dependence on system sizes and a linear dependence on the number of $\bold{k}$-points. 
The achieved efficiency stems from the integration of the $\bold{q}$-dependent ISDF procedure with the space-group symmetries inherent in crystalline systems.

By leveraging the complete separability of both orbital and $\bold{k}$-point indices in the ISDF expansion for a product basis, the evaluation of $GW$ polarizability and self-energy can be efficiently conducted in space-time coordinates, resembling equations in the conventional space-time formalism. 
Moreover, as the dimension of spatial coordinates corresponds to the number of interpolating points rather than the size of a uniform real-space discretization, the prefactors of ISDF-$GW$ and ISDF-cRPA are, in principle, much smaller compared to the space-time formalism, especially in the presence of hard pseudopotentials. 
Simultaneously, the incorporation of space-group symmetries within ISDF further reduces the size of the auxiliary basis, significantly simplifying computational and memory requirements for bosonic quantities such as polarizability and screened interaction.

The accuracy of ISDF-$GW$ and ISDF-cRPA is validated through an analysis of the convergence of the ISDF auxiliary basis and a benchmark on the band gaps of single-shot and fully self-consistent variants of $GW$. 
Finally, the capability of symmetry-adapted ISDF in treating large-scale systems is demonstrated through the construction of downfolded MB Hamiltonians for $\mathrm{C_{B}C_{N}}$ defects in hBN supercells. 
Our work demonstrates the efficiency and general applicability of ISDF in the context of large-scale many-body calculations with $\bold{k}$-point sampling beyond DFT. 

\section*{Supporting Information}

Basis set and finite-size convergence for the data presented in Tables.~\ref{tab:g0w0_band_gaps_ccecp} and~\ref{tab:scgw_band_gaps_ccecp}; derivations of the downfolded Hamiltonian $\hat{H}^{\mathrm{PBE+cRPA}}_{\mathrm{eff}}$ and $\hat{H}^{G_{0}W_{0}+\mathrm{cRPA}}_{\mathrm{eff}}$. 

\begin{acknowledgments}

We thank Cyrus Dreyer and Woncheol Lee for helpful discussion and providing reference data for $\mathrm{C_{B}C_{N}}$ defect systems. 
The Flatiron Institute is a division of the Simons Foundation.

\end{acknowledgments}

\bibliographystyle{apsrev4-2}
\bibliography{./refs}

\end{document}


\newpage

\section{Basis set convergence}
\begin{figure}[tbh!]
\begin{center}
\includegraphics[width=0.32\textwidth]{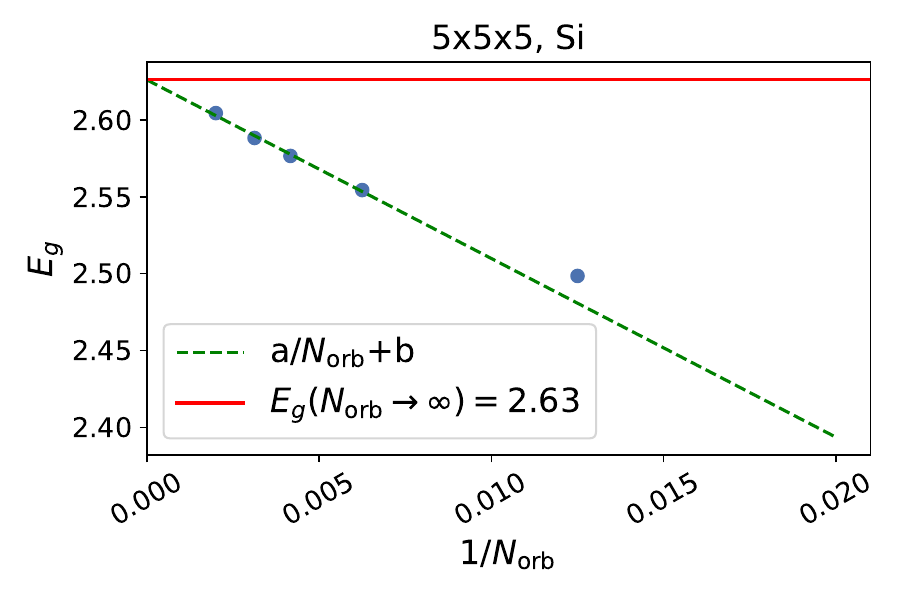} 
\includegraphics[width=0.32\textwidth]{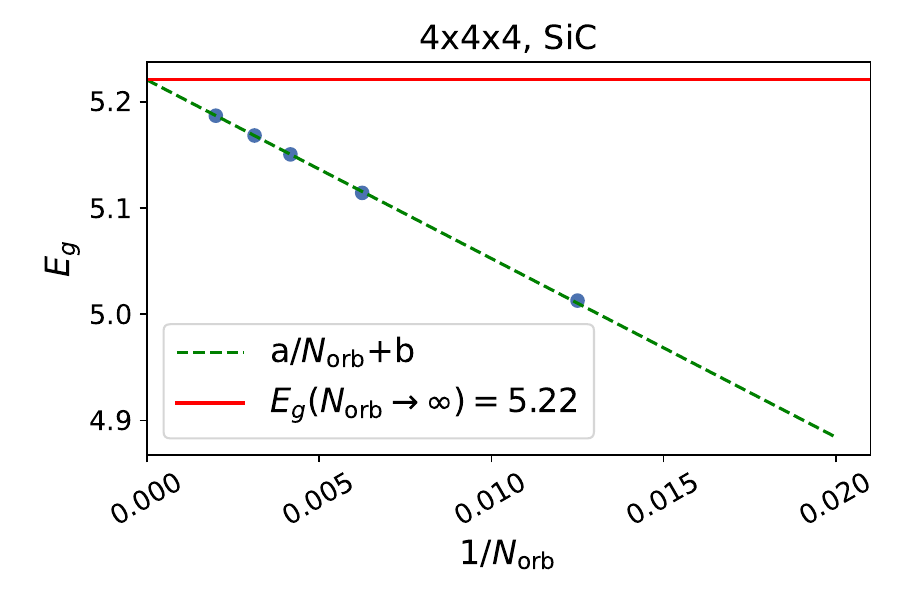} 
\includegraphics[width=0.32\textwidth]{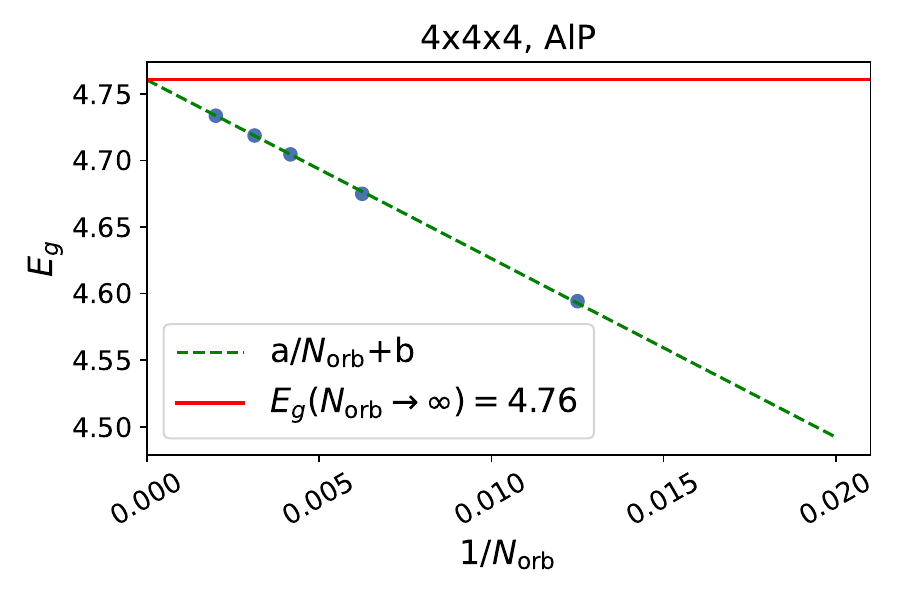} \\
\includegraphics[width=0.32\textwidth]{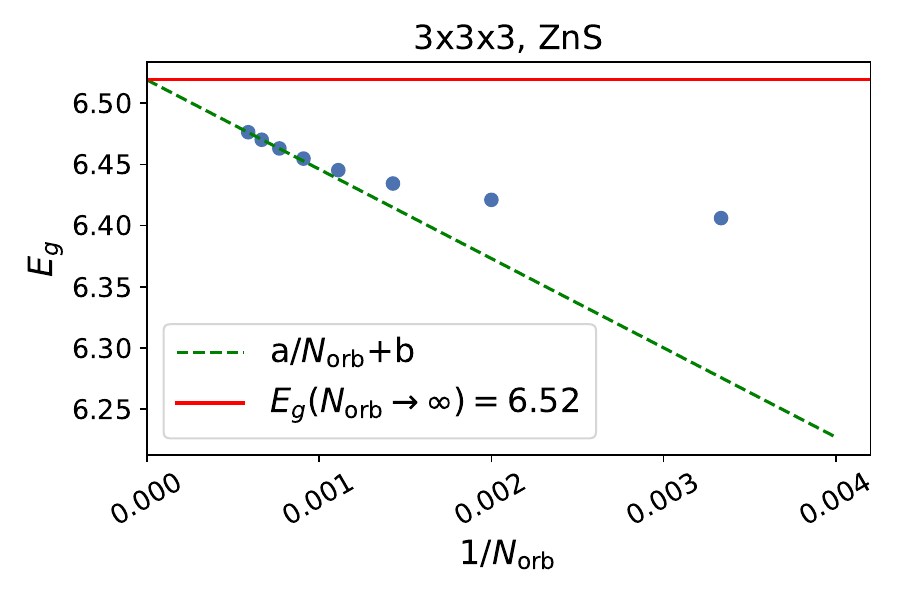} 
\includegraphics[width=0.32\textwidth]{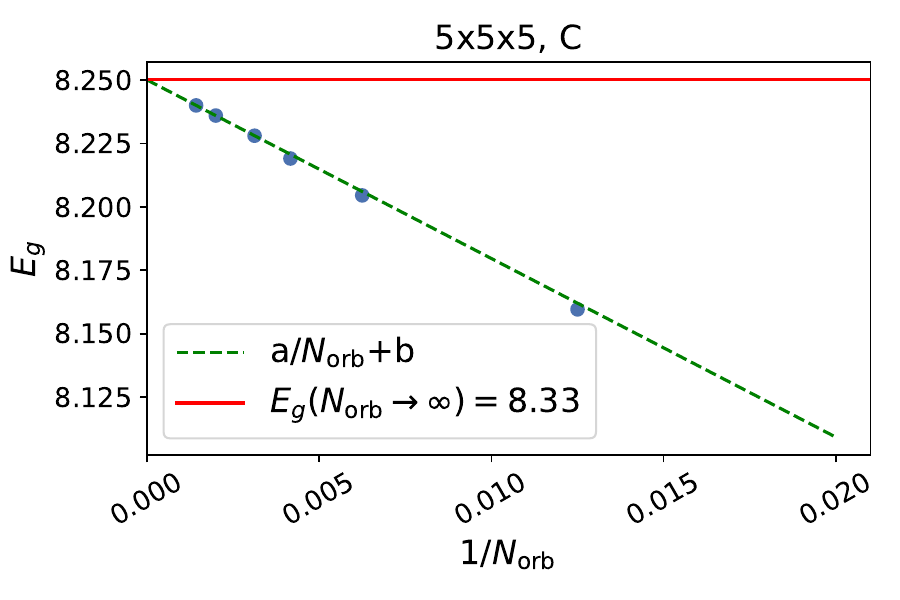} 
\includegraphics[width=0.32\textwidth]{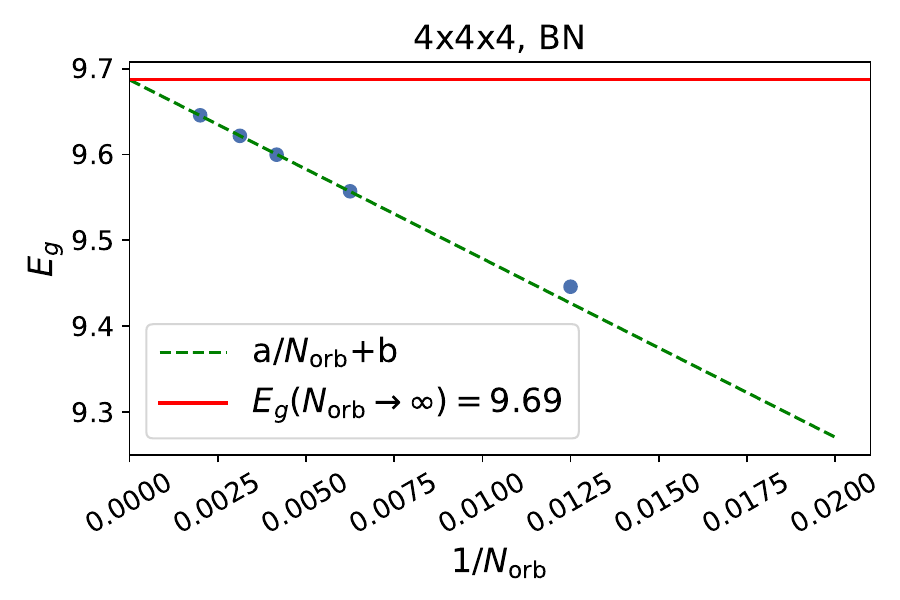} \\
\includegraphics[width=0.32\textwidth]{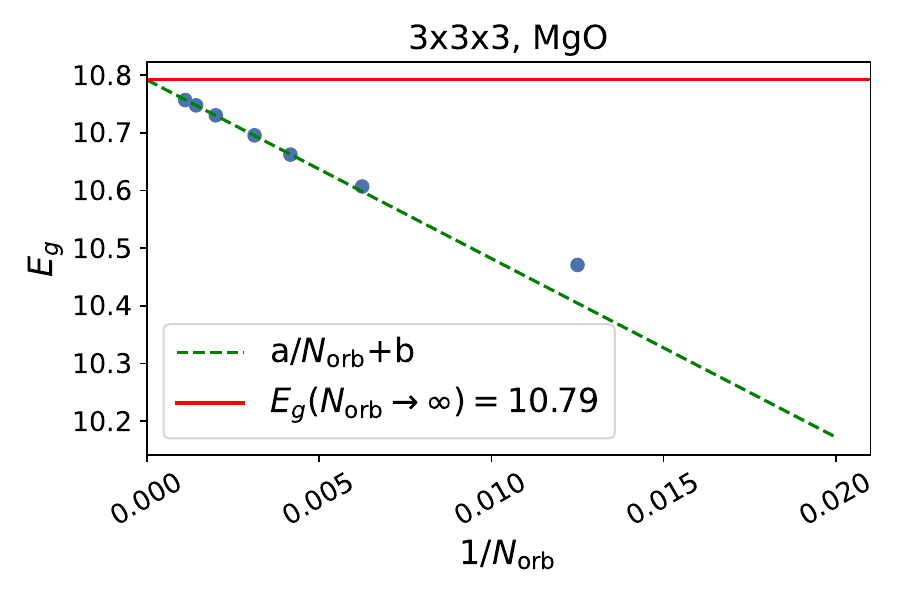} 
\includegraphics[width=0.32\textwidth]{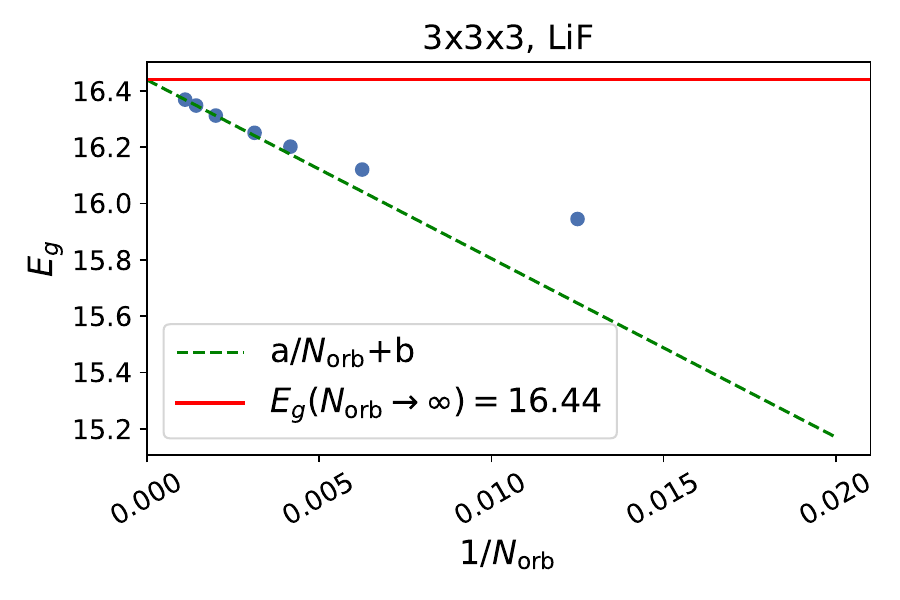} \\
\caption{Basis convergence of ISDF-$G_{0}W_{0}$ band gaps. }
\label{fig:basis_conv_g0w0}
\end{center}
\end{figure}
\begin{figure}[tbh!]
\begin{center}
\includegraphics[width=0.32\textwidth]{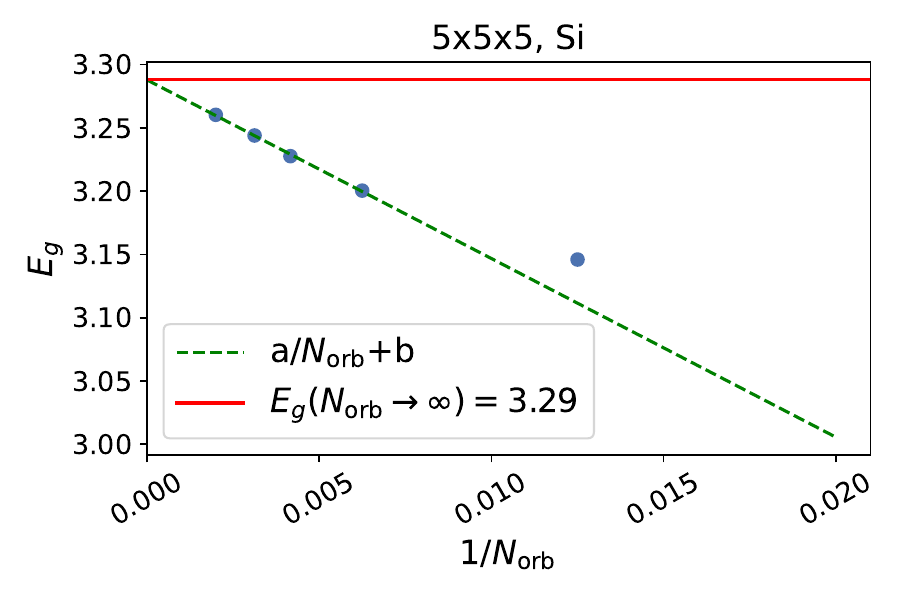} 
\includegraphics[width=0.32\textwidth]{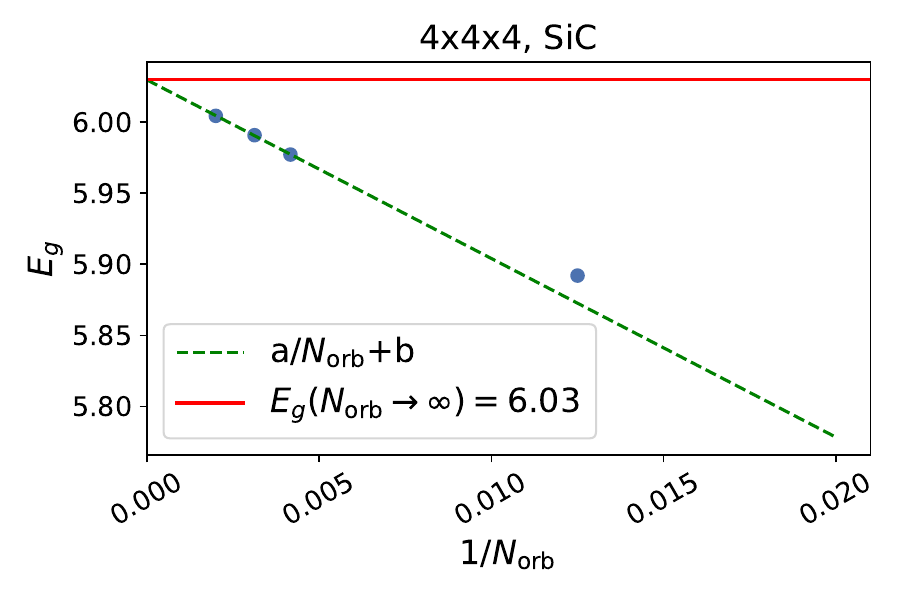} 
\includegraphics[width=0.32\textwidth]{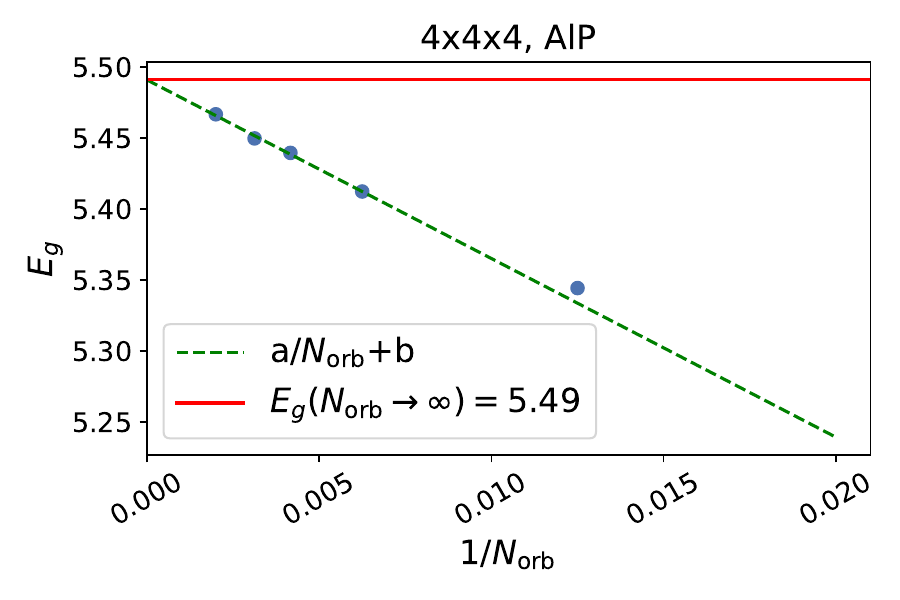} \\
\includegraphics[width=0.32\textwidth]{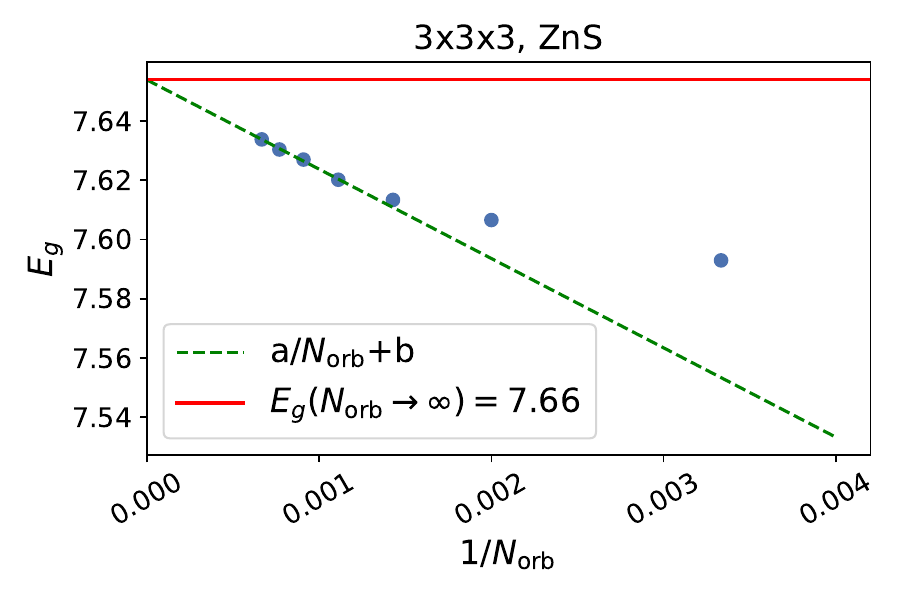} 
\includegraphics[width=0.32\textwidth]{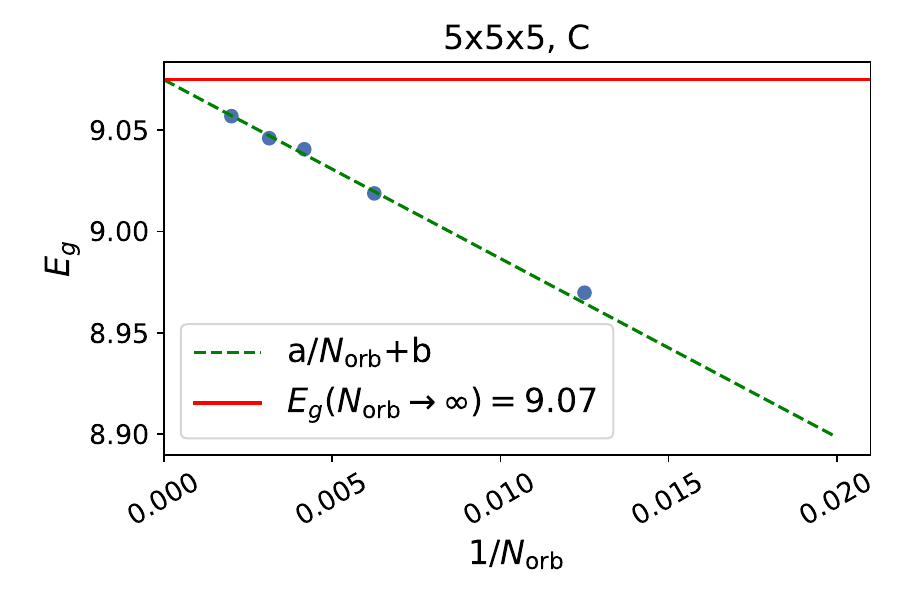} 
\includegraphics[width=0.32\textwidth]{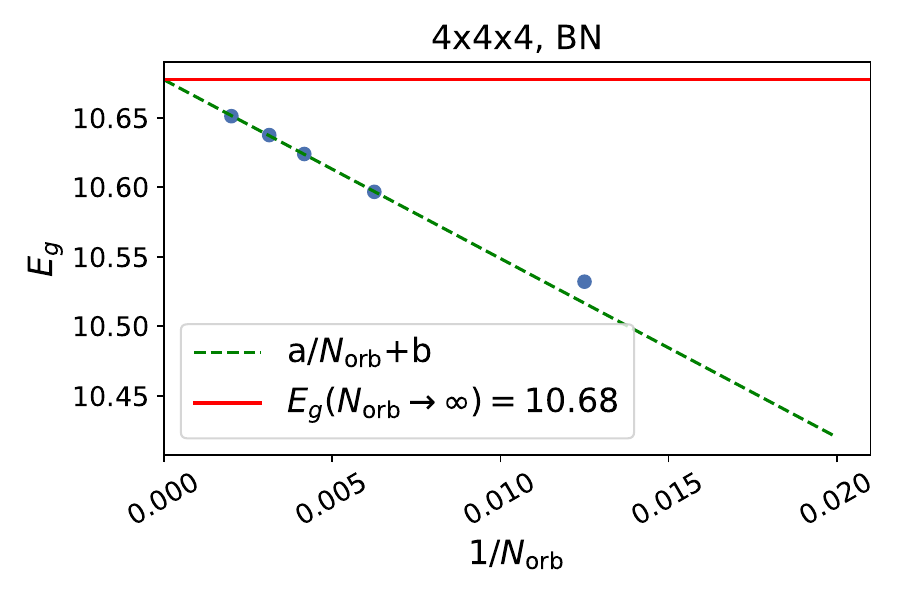} \\
\includegraphics[width=0.32\textwidth]{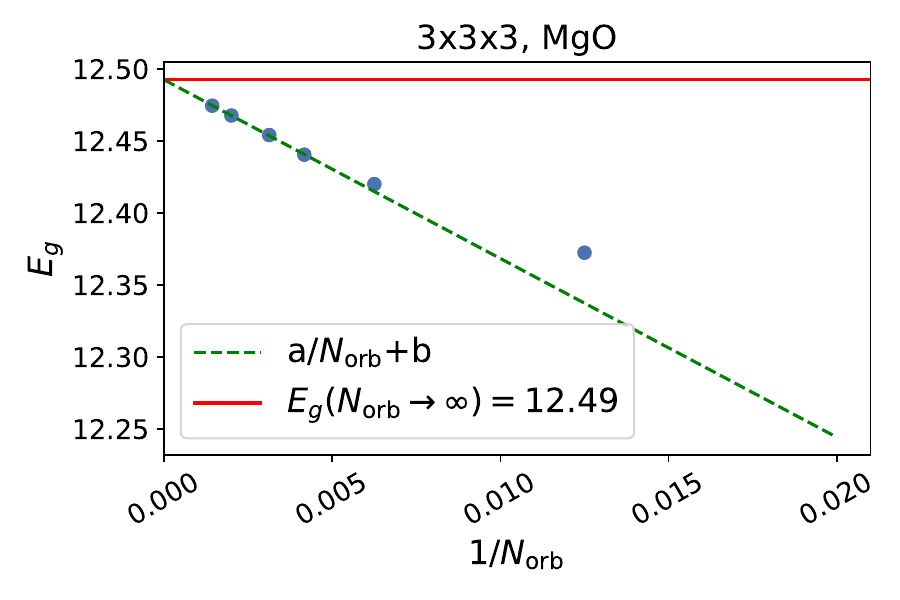} 
\includegraphics[width=0.32\textwidth]{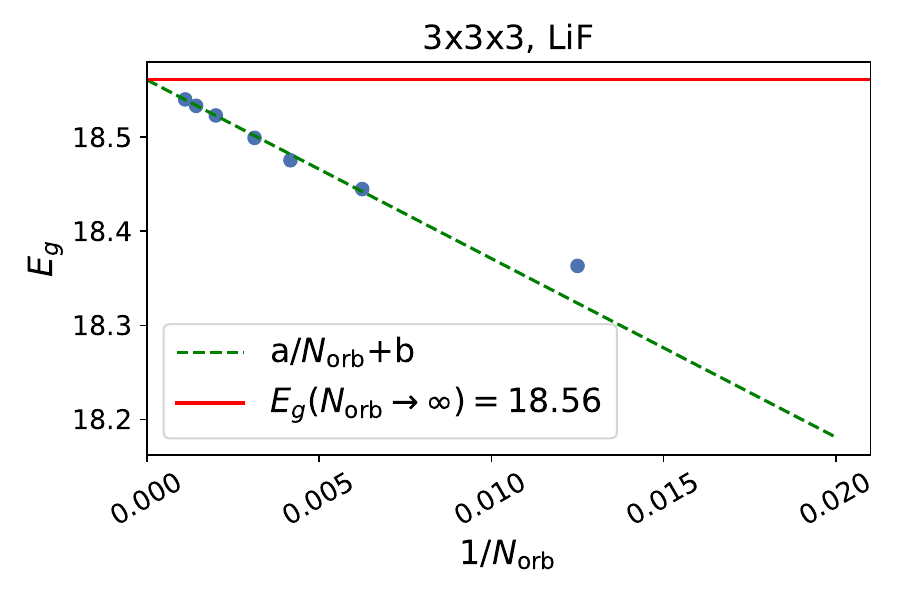} \\
\caption{Basis convergence of ISDF-sc$GW$ band gaps. }
\label{fig:basis_conv_scgw}
\end{center}
\end{figure}

In this section, we show the basis set convergence of the band gaps presented in Tables II and III in the main text. 
As depicted in Fig.\ref{fig:basis_conv_g0w0}, the band gap exhibits a consistent linear dependence on the number of basis functions $N_{\mathrm{orb}}$ when $N_{\mathrm{orb}}$ is sufficiently large, across all selected systems. 
Here, the basis set correction is determined by quantifying the deviation from the complete basis set limit, fitting the formula $E_{g}(N_{\mathrm{orb}}) = a/N_{\mathrm{orb}} + b$. 
For all selected systems, these corrections consistently remain below 0.15 eV. Similar analyses for the band gaps calculated using ISDF-sc$GW$ are presented in Fig.~\ref{fig:basis_conv_scgw}. 

\clearpage

\section{Convergence to the thermodynamic limit}
\begin{figure}[tbh!]
\begin{center}
\includegraphics[width=0.32\textwidth]{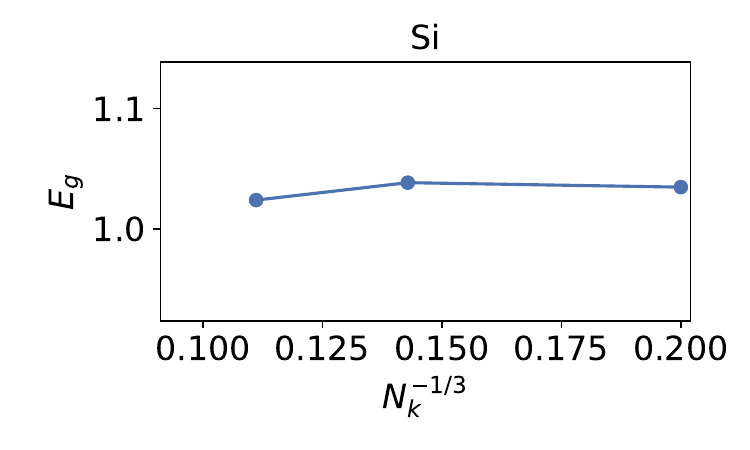} 
\includegraphics[width=0.32\textwidth]{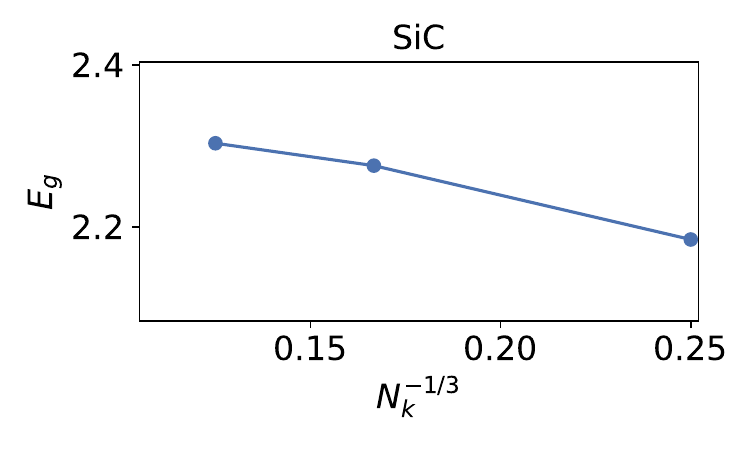} 
\includegraphics[width=0.32\textwidth]{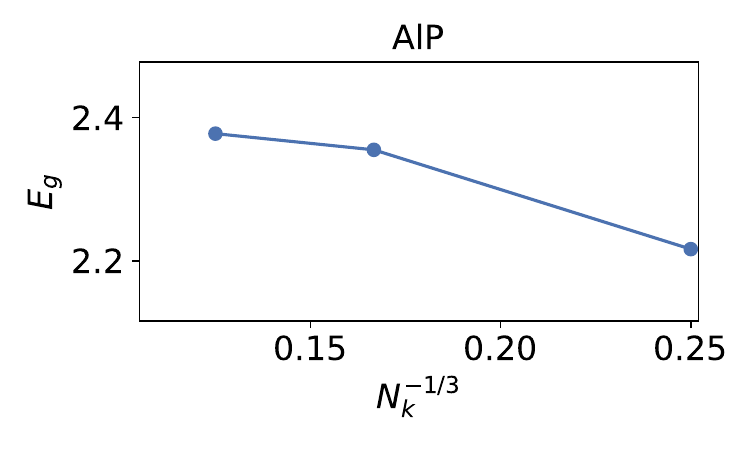} \\
\includegraphics[width=0.32\textwidth]{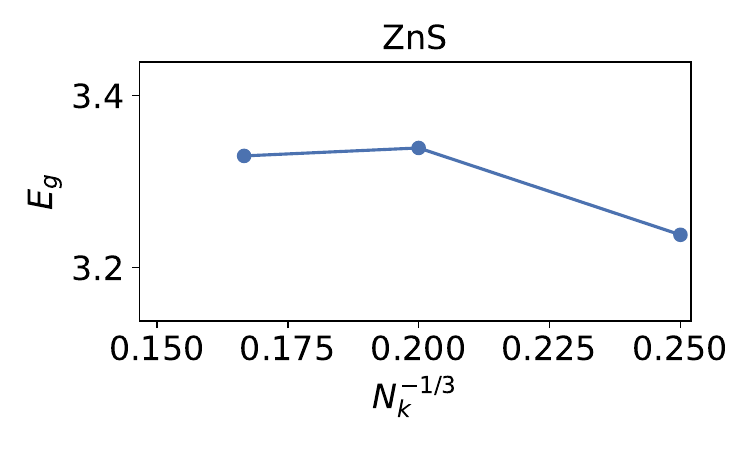} 
\includegraphics[width=0.32\textwidth]{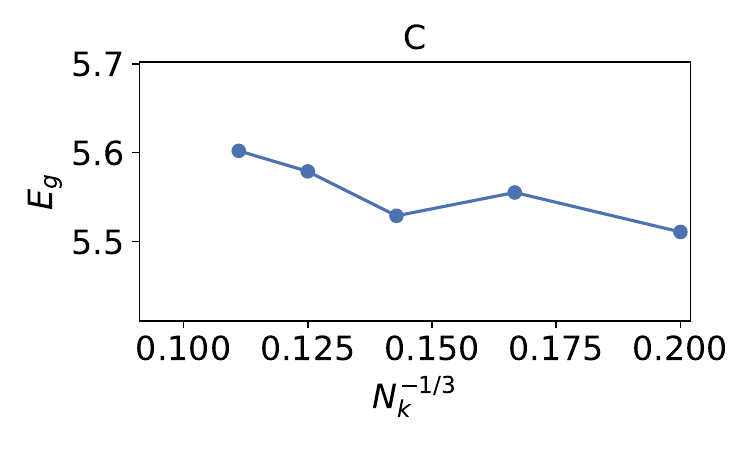} 
\includegraphics[width=0.32\textwidth]{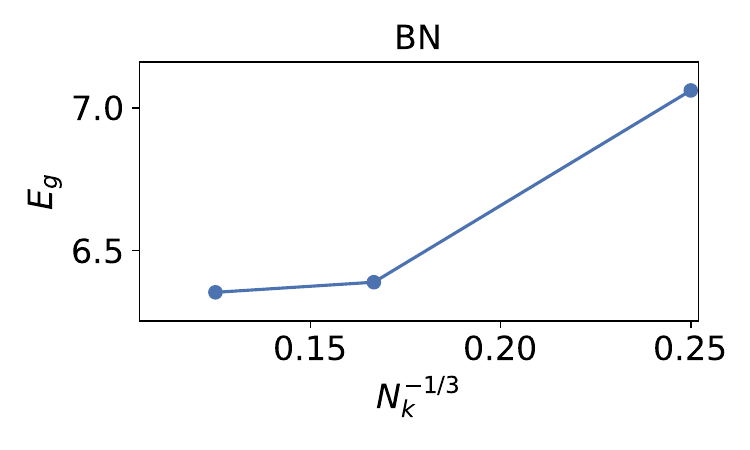} \\
\includegraphics[width=0.32\textwidth]{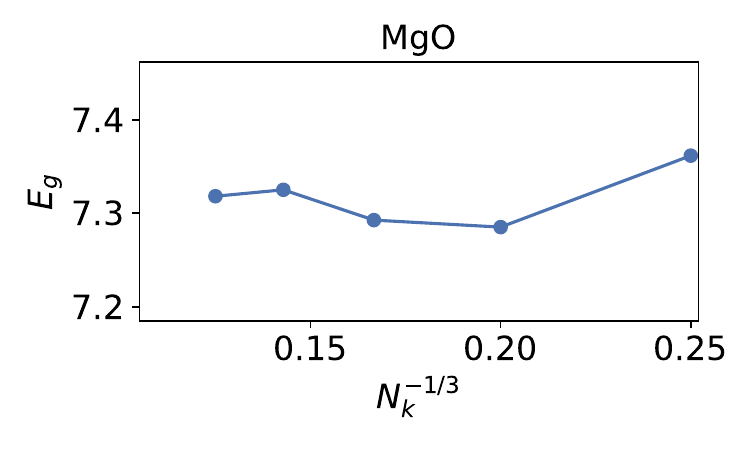} 
\includegraphics[width=0.32\textwidth]{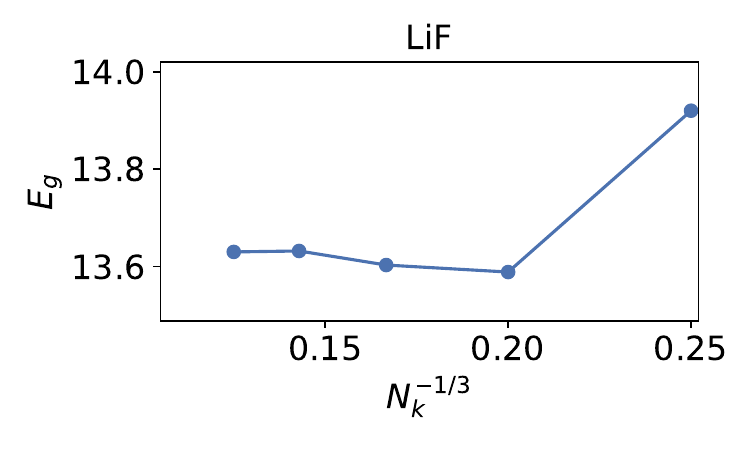} \\
\caption{Convergence of ISDF-$G_{0}W_{0}$ band gaps with the total number of $\textbf{k}$-points.}
\label{fig:k_conv_g0w0}
\end{center}
\end{figure}

\begin{figure}[tbh!]
\begin{center}
\includegraphics[width=0.32\textwidth]{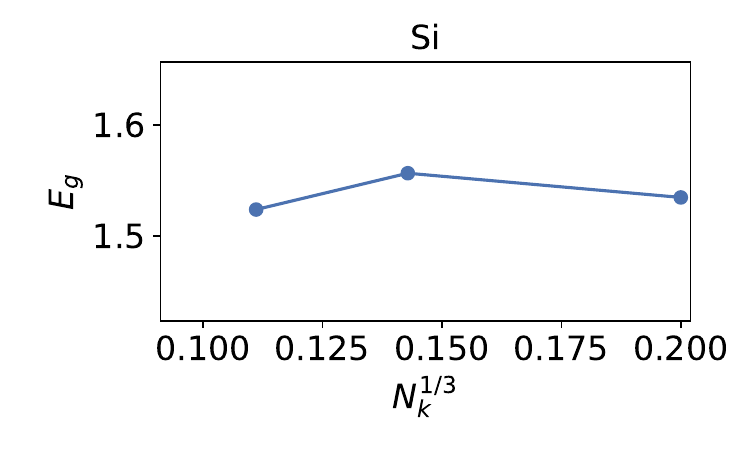} 
\includegraphics[width=0.32\textwidth]{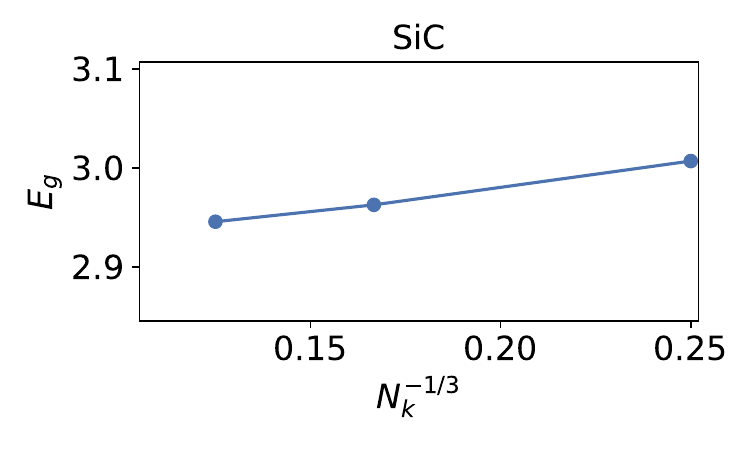} 
\includegraphics[width=0.32\textwidth]{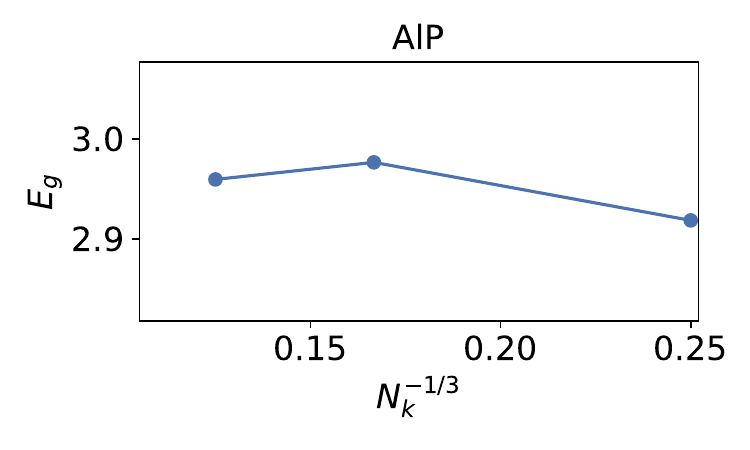} \\
\includegraphics[width=0.32\textwidth]{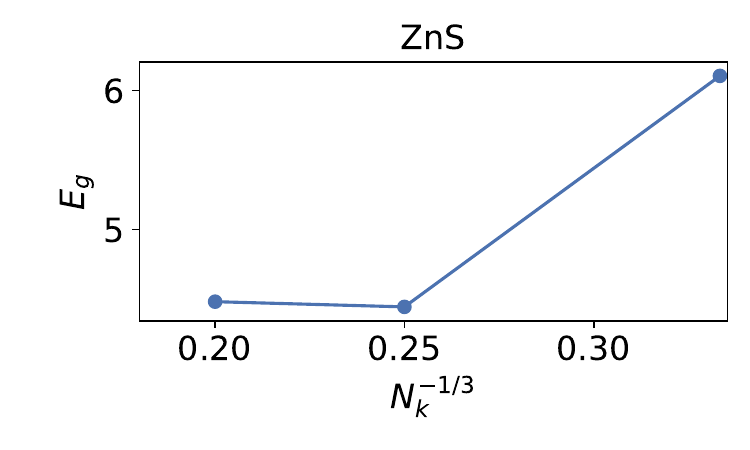} 
\includegraphics[width=0.32\textwidth]{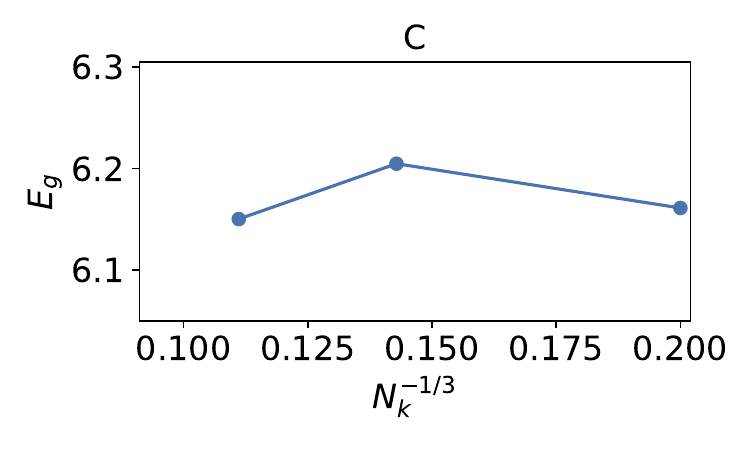} 
\includegraphics[width=0.32\textwidth]{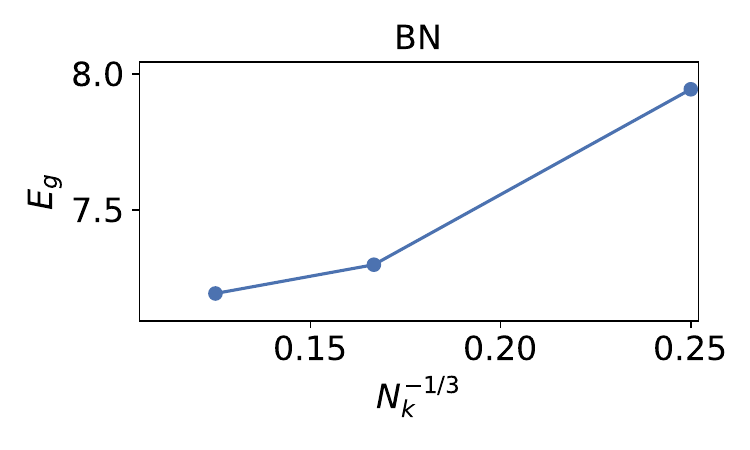} \\
\includegraphics[width=0.32\textwidth]{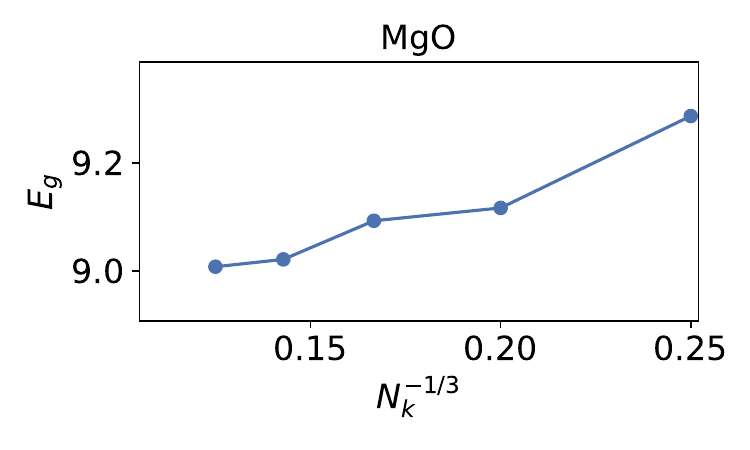} 
\includegraphics[width=0.32\textwidth]{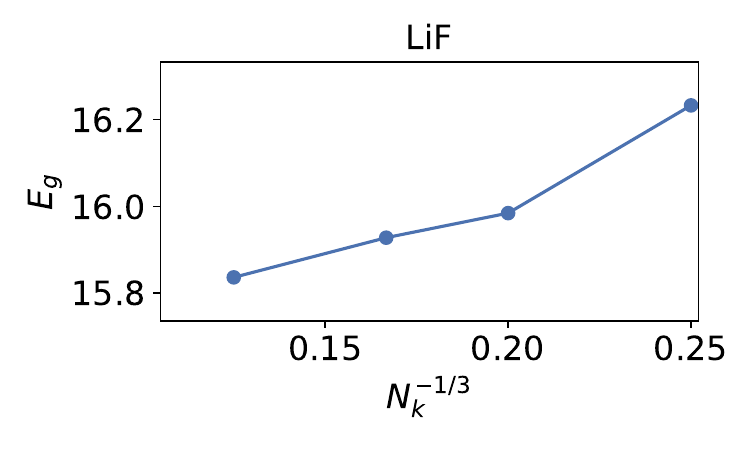} \\
\caption{Convergence of ISDF-sc$GW$ band gaps with the total number of $\textbf{k}$-points.}
\label{fig:k_conv_scgw}
\end{center}
\end{figure}

In this section, we present the convergence to the thermodynamic limit of the band gaps presented in Tables II and III in the main text. Figures~\ref{fig:k_conv_g0w0} and~\ref{fig:k_conv_scgw} respectively show the band gaps calculated using ISDF-$G_{0}W_{0}$ and ISDF-sc$GW$, as a function of the number of $\textbf{k}$-points sampled in the Brillouin zone. 

We add the finite-size corrections to the head of the $GW$ self-energy using the correction scheme proposed by Gygi and Baldereschi~\cite{Gygi_corrections1986}. It is important to note that the wings finite-size corrections are not considered in the present work. We anticipate that including such corrections will further enhance convergence, especially for ISDF-sc$GW$, where the off-diagonal terms of the self-energy play a crucial role.

\clearpage

\section{Effective Hamiltonian for \texorpdfstring{$\mathrm{C_{B}C_{N}}$  \label{app:eff_hamilt}}{CBCN}}
In this section, we outline the formula for the one-body and two-body components of the effective Hamiltonians ($\hat{H}^{G_{0}W_{0}\mathrm{+cRPA}}_{\mathrm{eff}}$ and $\hat{H}^{\mathrm{PBE+cRPA}}_{\mathrm{eff}}$) presented in Sec. VII.D in the main text. 
Our formulations for $\hat{H}^{\mathrm{PBE+cRPA}}_{\mathrm{eff}}$ and $\hat{H}^{G_{0}W_{0}\mathrm{+cRPA}}_{\mathrm{eff}}$ closely follow the quantum embedding approach for defect systems elucidated in Ref.\citenum{PRB_Muechler2022} and Ref.\citenum{QDET_Sheng2022}, respectively. 

Starting with a PBE non-interacting Green's function: 
\begin{align}
(G_{0})^{\textbf{k}}_{ij}(i\omega_{n}) = \frac{\delta_{ij}}{i\omega_{n} + \mu - \epsilon^{\mathrm{KS}}_{i\textbf{k}}} 
\end{align}
where $\epsilon^{\mathrm{KS}}_{i\textbf{k}}$ denotes the band energy for the Kohn-Sham (KS) orbital $\phi^{\textbf{k}}_{i}(\textbf{r})$, we aim to construct an effective Hamiltonian written as: 
\begin{align}
\hat{H}_{\mathrm{eff}} = \sum_{\alpha\beta}(t^{\mathrm{eff}}_{\alpha\beta}-\mu\delta_{\alpha\beta})\hat{c}^{\dag}_{\alpha}\hat{c}_{\beta} + \frac{1}{2}\sum_{\alpha\beta\gamma\delta}v^{\mathrm{eff}}_{\alpha\beta\gamma\delta}\hat{c}^{\dag}_{\alpha}\hat{c}^{\dag}_{\gamma}\hat{c}_{\delta}\hat{c}_{\beta}.  
\label{eq:eff_hamilt}
\end{align}
Here, $\textbf{t}^{\mathrm{eff}}$ and $\textbf{v}^{\mathrm{eff}}$, respectively, denote the effective one-body and two-body Hamiltonians, $\mu$ is the chemical potential and the Greek letters denote the local Wannier functions $w_{\alpha}(\textbf{r})$ for the $\mathrm{C_{B}C_{N}}$ defects states: 
\begin{align}
w_{\alpha}(\textbf{r}) = \frac{1}{N_{k}}\sum_{\textbf{k}}\sum_{i}C^{\textbf{k}*}_{\alpha i}\phi^{\textbf{k}}_{i}(\textbf{r})e^{-i\textbf{kR}}\big|_{\textbf{R=0}}, 
\end{align}
where $C^{\textbf{k}}_{\alpha i} = \langle w^{\textbf{k}}_{\alpha}| \phi^{\textbf{k}}_{i}\rangle$ characterizes the orbital transformation between the Wannier $w^{\textbf{k}}_{\alpha}(\textbf{r})$ and KS $\phi^{\textbf{k}}_{i}(\textbf{r})$ orbitals, and $\textbf{R}$ is a translational vector of the system. 

\subsection{Two-body Hamiltonian}
Starting with a PBE non-interacting Green's function $(G_{0})^{\textbf{k}}_{ij}(i\omega_{n})$, we first calculate the local cRPA screened interactions $U_{\alpha\beta\gamma\delta}(i\Omega_{n})[\textbf{G}^{\textbf{k}}_{0}]$ on the Matsubara frequency axis, following the methodology outlined in Sec. V.C in the main text. 
Subsequently, the effective two-body Hamiltonians for $H^{\mathrm{PBE+cRPA}}_{\mathrm{eff}}$ and $H^{G_{0}W_{0}\mathrm{+cRPA}}_{\mathrm{eff}}$ are derived by applying a static approximation to $\mathbf{U}(i\Omega_{n})$ at $i\Omega_{n}=0$\cite{PRB_Muechler2022,QDET_Sheng2022}: 
\begin{align}
v^{\mathrm{eff}}_{\alpha\beta\gamma\delta} = U_{\alpha\beta\gamma\delta}(i\Omega_{n}=0).   
\end{align}

\subsection{One-body Hamiltonian}
\subsubsection{\texorpdfstring{$H^{\mathrm{PBE+cRPA}}_{\mathrm{eff}}$}{HPBEcRPAeff}\label{subsubsec:PBE_cRPA}}
The effective one-body Hamiltonian $\textbf{t}^{\mathrm{eff}}$ for $H^{\mathrm{PBE+cRPA}}_{\mathrm{eff}}$ is expressed as follows~\cite{PRB_Muechler2022}:  
\begin{subequations}
\begin{align}
t^{\mathrm{eff}}_{\alpha\beta} &= (H_{0})_{\alpha\beta} + J_{\alpha\beta} + V^{\mathrm{xc}}_{\alpha\beta} - (H_{\mathrm{DC}})_{\alpha\beta} \\
&= (H^{\mathrm{PBE}})_{\alpha\beta} - (H_{\mathrm{DC}})_{\alpha\beta}. 
\end{align}
\end{subequations}
Here, $\textbf{H}^{\mathrm{PBE}}$ is the PBE Hamiltonian comprising the non-interacting Hamiltonian $\textbf{H}_{0}$, the Coulomb (Hartree) potential $\textbf{J}$ and the exchange-correlation potential $\textbf{V}^{\mathrm{xc}}$. 
The double counting (DC) term $\textbf{H}_{\mathrm{DC}}$ is subtracted to account for correlations considered by both PBE and exact diagonalization (ED). 
In practice, $\textbf{H}^{\mathrm{PBE}}$ can be directly evaluated from the KS band energies: 
\begin{align}
(H^{\mathrm{PBE}})_{\alpha\beta} = \frac{1}{N_{k}}\sum_{\textbf{k}} \sum_{ij} C^{\textbf{k}}_{\alpha i} \epsilon^{\textbf{k}}_{i}  C^{\textbf{k}*}_{\beta i}.  
\end{align}

Applying the approach detailed in Refs.\citenum{Bockstedte2018,PRB_Muechler2022}, we exclude the contributions from the exchange-correlation potential for double counting (DC) and approximate the DC term as the Hartree potential, computed using the effective interaction $\textbf{v}{\mathrm{eff}}$:
\begin{align}
(H_{\mathrm{DC}})_{\alpha\beta} \approx (J_{\mathrm{DC}})_{\alpha\beta}[\boldsymbol{\rho}_{0}, \textbf{v}^{\mathrm{eff}}]  = 2\sum_{\gamma\delta} (\rho_{0})_{\gamma\delta} v^{\mathrm{eff}}_{\alpha\beta\delta\gamma},
\end{align}
where $(\rho_{0})_{\gamma\delta}$ is the PBE density in the Wannier basis, and the factor of two takes into account the spin degrees of freedom. 

\subsubsection{\texorpdfstring{$H^{G_{0}W_{0}\mathrm{+cRPA}}_{\mathrm{eff}}$}{HG0w0cRPAeff}}
Starting with the static (also known as the exact exchange) and dynamic components of the $GW$ self-energy, as evaluated using the PBE Green's function $(G_{0})^{\textbf{k}}_{ij}(i\omega_{n})$: 
\begin{align}
\Sigma^{\textbf{k}}_{ij}(i\omega_{n}) = K^{\textbf{k}}_{ij} + \tilde{\Sigma}^{\textbf{k}}_{ij}(i\omega_{n}), 
\end{align}
the effective one-body Hamiltonian $\textbf{t}^{\mathrm{eff}}$ can be expressed as~\cite{QDET_Sheng2022}: 
\begin{align}
t^{\mathrm{eff}}_{\alpha\beta} = (H_{0})_{\alpha\beta} + J_{\alpha\beta} + K_{\alpha\beta} + V^{G_{0}W_{0}}_{\alpha\beta} - (H_{\mathrm{DC}})_{\alpha\beta}. 
\label{eq:t_g0w0}
\end{align}
Here, $\textbf{H}_{0}$ and $\textbf{J}$ are defined in Section~\ref{subsubsec:PBE_cRPA}. 
The matrix elements $\textbf{K}$ and $\textbf{V}^{G_{0}W_{0}}$ correspond to the contributions from the static and dynamic parts of the $GW$ self-energy, respectively. 
Finally, $\textbf{H}_{\mathrm{DC}}$ represents the corresponding DC terms. 

The first three terms in Eq.~\ref{eq:t_g0w0} can be straightforwardly obtained through the basis transformation from the KS basis to the Wannier orbitals:
\begin{align}
&(H_{0})_{\alpha\beta} = \frac{1}{N_{k}} \sum_{\textbf{k}} \sum_{ij} C^{\textbf{k}}_{\alpha i} (H^{\textbf{k}}_{0})_{ij} C^{\textbf{k}*}_{\beta j}, \\
&J_{\alpha\beta}[\boldsymbol{\rho^{\textbf{k}}}_{0}] = \frac{1}{N_{k}} \sum_{\textbf{k}} \sum_{ij} C^{\textbf{k}}_{\alpha i} J^{\textbf{k}}_{ij}[\boldsymbol{\rho^{\textbf{k}}}_{0}] C^{\textbf{k}*}_{\beta j} \\
&K_{\alpha\beta}[\boldsymbol{\rho^{\textbf{k}}}_{0}] = \frac{1}{N_{k}} \sum_{\textbf{k}} \sum_{ij} C^{\textbf{k}}_{\alpha i} K^{\textbf{k}}_{ij}[\boldsymbol{\rho^{\textbf{k}}}_{0}] C^{\textbf{k}*}_{\beta j}. 
\end{align}
Here, it is crucial to emphasize that, in the context of $G_{0}W_{0}$, the Coulomb ($J^{\textbf{k}}_{ij}$) and exchange ($K^{\textbf{k}}_{ij}$) potentials in the KS basis are functionals of the PBE non-interacting density $\boldsymbol{\rho^{\textbf{k}}}_{0}$, rather than the density from $G_{0}W_{0}$. 

The effective potential $\textbf{V}^{G_{0}W_{0}}$ arising from the dynamic part of the $GW$ self-energy necessitates an additional approximation for its frequency dependence. 
In the present work, we apply the widely used static approximation proposed by Kotani \emph{et al.}~\cite{QPGW_Kotani_PRL2004} in the KS basis: 
\begin{align}
&(V^{G_{0}W_{0}})^{\textbf{k}}_{ij} = \frac{1}{4} \big[  \tilde{\Sigma}^{\textbf{k}}_{ij}(\epsilon^{G_{0}W_{0}}_{i\textbf{k}}-\mu) + \tilde{\Sigma}^{\textbf{k}*}_{ji}(\epsilon^{G_{0}W_{0}}_{i\textbf{k}}-\mu) + \tilde{\Sigma}^{\textbf{k}}_{ij}(\epsilon^{G_{0}W_{0}}_{j\textbf{k}}-\mu) + \tilde{\Sigma}^{\textbf{k}*}_{ji}(\epsilon^{G_{0}W_{0}}_{j\textbf{k}}-\mu)\big], 
\end{align}
and subsequently, we downfold the matrix elements to the local Wannier basis: 
\begin{align}
V^{G_{0}W_{0}}_{\alpha\beta} = \frac{1}{N_{k}} \sum_{\textbf{k}} \sum_{ij} C^{\textbf{k}}_{\alpha i} (V^{G_{0}W_{0}})^{\textbf{k}}_{ij} C^{\textbf{k}*}_{\beta j}.  
\end{align}
Here, $\epsilon^{G_{0}W_{0}}_{i\textbf{k}}$ represents the $i$-th $G_{0}W_{0}$ quasiparticle energies, and it is important to note that these should not be confused with the KS energies $\epsilon^{\mathrm{KS}}_{i\textbf{k}}$. 

Conceptually, the DC term $\textbf{H}^{\mathrm{DC}}$ captures to the correlations at the $GW$ level within the local effective Hamiltonian. 
Since we do not incorporate frequency-dependent Coulomb interactions in our effective Hamiltonian, $\textbf{H}^{\mathrm{DC}}$ can be precisely formulated using the local PBE Green's function $\textbf{G}_{0}$ and the effective interactions $\textbf{v}_{\mathrm{eff}}$ from $\hat{H}_{\mathrm{eff}}$ within the framework of the Baym-Kadanoff formulation~\cite{Baym61,Baym62}: 
\begin{align}
(H_{\mathrm{DC}})_{\alpha\beta}[\boldsymbol{\rho}_{0}, \textbf{v}^{\mathrm{eff}}] = (J_{\mathrm{DC}})_{\alpha\beta}[\boldsymbol{\rho}_{0}, \textbf{v}^{\mathrm{eff}}] + (K_{\mathrm{DC}})_{\alpha\beta}[\boldsymbol{\rho}_{0}, \textbf{v}^{\mathrm{eff}}] + (V^{G_{0}W_{0}}_{\mathrm{DC}})_{\alpha\beta}[\boldsymbol{\rho}_{0}, \textbf{v}^{\mathrm{eff}}]
\label{eq:dc_g0w0}
\end{align}
where $\textbf{J}_{\mathrm{DC}}$ represents the local Coulomb potential, and $\textbf{K}_{\mathrm{DC}}$ and $\textbf{V}^{G_{0}W_{0}}_{\mathrm{DC}}$ collectively account for the $GW$ self-energy diagrams within the local effective Hamiltonian.  

The first two terms in Eq.~\ref{eq:dc_g0w0} collectively represent the Hartree-Fock (HF) potential: 
\begin{align}
(J_{\mathrm{DC}})_{\alpha\beta}[\boldsymbol{\rho}_{0}, \textbf{v}^{\mathrm{eff}}] + (K_{\mathrm{DC}})_{\alpha\beta}[\boldsymbol{\rho}_{0}, \textbf{v}^{\mathrm{eff}}] = \sum_{\gamma\delta} (\rho_{0})_{\gamma\delta} \Big[ 2v^{\mathrm{eff}}_{\alpha\beta\delta\gamma} - v^{\mathrm{eff}}_{\alpha\gamma\delta\beta} \Big],  
\end{align}
where $\boldsymbol{\rho}_{0}$ denotes the local PBE density in the Wannier basis and $\textbf{v}^{\mathrm{eff}}$ represents the effective interactions as given in Eq.~\ref{eq:eff_hamilt}. 

$\textbf{V}^{G_{0}W_{0}}_{\mathrm{DC}}$ is derived from the dynamic part of the $GW$ self-energy $(\tilde{\Sigma}^{G_{0}W_{0}}_{\mathrm{DC}})_{\alpha\beta}(\tau)$ for $\hat{H}_{\mathrm{eff}}$: 
\begin{align}
&(\tilde{\Sigma}^{G_{0}W_{0}}_{\mathrm{DC}})_{\alpha\beta}(\tau)[\textbf{G}_{0}, \textbf{v}^{\mathrm{eff}}] = -\sum_{\delta\gamma}(G_{0})_{\gamma\delta}(\tau)\tilde{W}^{\mathrm{eff}}_{\alpha\gamma\delta\beta}(\tau) \\ 
&\tilde{W}^{\mathrm{eff}}_{\alpha\gamma\delta\beta}(i\Omega_{n})[\textbf{G}_{0}, \textbf{v}^{\mathrm{eff}}] = \Big[ \textbf{I} - \textbf{v}^{\mathrm{eff}}\boldsymbol{\Pi}^{\mathrm{eff}}(i\Omega_{n})\Big]\textbf{v}^{\mathrm{eff}} - \textbf{v}^{\mathrm{eff}} \\ 
&\Pi^{\mathrm{eff}}_{\alpha\beta\gamma\delta}(\tau)[\textbf{G}_{0}] = (G_{0})_{\beta\gamma}(\tau)(G_{0})_{\delta\alpha}(-\tau).  
\end{align}
Here, $\boldsymbol{\Pi}^{\mathrm{eff}}$ and $\tilde{\textbf{W}}^{\mathrm{eff}}$ correspond to the effective polarizability and the effective dynamic screened interactions. 
It is crucial to emphasize that both $\boldsymbol{\Pi}^{\mathrm{eff}}$ and $\tilde{\textbf{W}}^{\mathrm{eff}}$ are functionals of $\textbf{G}_{0}$ and $\textbf{v}^{\mathrm{eff}}$. 
Finally, a consistent treatment for the frequency dependence of $(\tilde{\Sigma}^{G_{0}W_{0}}_{\mathrm{DC}})_{\alpha\beta}(i\omega_{n})$ is applied: 
\begin{align}
&(\tilde{\Sigma}^{G_{0}W_{0}}_{\mathrm{DC}})^{\textbf{k}}_{ij}(i\omega_{n}) = \sum_{\alpha\beta} C^{\textbf{k}*}_{\alpha i} (\tilde{\Sigma}^{G_{0}W_{0}}_{\mathrm{DC}})^{\textbf{k}}_{\alpha\beta}(i\omega_{n})C^{\textbf{k}}_{\beta j}\\
&(V^{G_{0}W_{0}}_{\mathrm{DC}})^{\textbf{k}}_{ij} = \frac{1}{4} \big[  (\tilde{\Sigma}^{G_{0}W_{0}}_{\mathrm{DC}})^{\textbf{k}}_{ij}(\epsilon^{G_{0}W_{0}}_{i\textbf{k}}-\mu) + (\tilde{\Sigma}^{G_{0}W_{0}}_{\mathrm{DC}})^{\textbf{k}*}_{ji}(\epsilon^{G_{0}W_{0}}_{i\textbf{k}}-\mu) \\
& \ \ \ \ \ \ \ \ \ \ \ \ \ \ \ \ \ \ \ \ \ \ \ \ \ \ \ \ \ \ \ \ + (\tilde{\Sigma}^{G_{0}W_{0}}_{\mathrm{DC}})^{\textbf{k}}_{ij}(\epsilon^{G_{0}W_{0}}_{j\textbf{k}}-\mu) + (\tilde{\Sigma}^{G_{0}W_{0}}_{\mathrm{DC}})^{\textbf{k}*}_{ji}(\epsilon^{G_{0}W_{0}}_{j\textbf{k}}-\mu)\big] \nonumber\\
&(V^{G_{0}W_{0}}_{\mathrm{DC}})_{\alpha\beta} = \frac{1}{N_{k}} \sum_{\textbf{k}} \sum_{ij} C^{\textbf{k}}_{\alpha i} (V^{G_{0}W_{0}}_{\mathrm{DC}})^{\textbf{k}}_{ij} C^{\textbf{k}*}_{\beta j}. 
\end{align}
Here, the static approximation is executed in the KS basis on the real frequency axis, and then downfolded back to the Wannier basis.

\bibliography{../refs}